\documentclass[aps,reprint,nofootinbib,superscriptaddress]{revtex4-2}

\usepackage{amsmath,amssymb,amsthm,mathtools,mathrsfs,hyperref}

\usepackage{graphicx}
\graphicspath{{PaperPlots/}}

\begin{document}

\title{Efficient simulations of high-spin black holes with a new gauge}

\author{Yitian Chen}
\email{yc2377@cornell.edu}
\affiliation{Cornell Center for Astrophysics and Planetary Science, Cornell University, Ithaca, New York 14853, USA\looseness=-1}

\author{Nils Deppe}
\affiliation{Theoretical Astrophysics 350-17, California Institute of Technology, Pasadena, California 91125, USA}

\author{Lawrence E. Kidder}
\affiliation{Cornell Center for Astrophysics and Planetary Science, Cornell University, Ithaca, New York 14853, USA\looseness=-1}

\author{Saul A. Teukolsky}
\affiliation{Cornell Center for Astrophysics and Planetary Science, Cornell University, Ithaca, New York 14853, USA\looseness=-1}
\affiliation{Theoretical Astrophysics 350-17, California Institute of Technology, Pasadena, California 91125, USA}

\date{\today}

\begin{abstract}
We present a new choice of initial data for binary black hole simulations that
significantly improves the efficiency of high-spin simulations. We use spherical
Kerr-Schild coordinates, where the horizon of a rotating black hole is
spherical, for each black hole. The superposed spherical Kerr-Schild initial
data reduce the runtime by a factor of 2 compared to standard superposed
Kerr-Schild for an intermediate resolution spin-0.99 binary-black-hole
simulation. We also explore different variations of the gauge conditions imposed during the evolution, one of which produces an additional speed-up of 1.3.
\end{abstract}

\maketitle

\section{Introduction} \label{sec:intro}

The exciting era of gravitational wave astronomy started with the first
detection of a gravitational wave (GW) in 2015 by the Laser Interferometer
Gravitational-Wave Observatory (LIGO) \cite{1602.03837, 1602.03839, 1602.03838,
1602.03840}. The event, named GW150914, was generated by a binary-black-hole
(BBH) system. About 50 detections have been published so far \cite{1811.12907, 2010.14527} and most of them were emitted by BBHs, including GW190521 \cite{2009.01075}, the first detection of an intermediate-mass black hole (BH). 

GW detection requires accurate waveform templates to extract the astrophysical
signal from the noisy data. While a full analytic solution of BBH
evolution is not known, analytical approximations like the post-Newtonian method
can generate approximate BBH waveforms \cite{1310.1528}.  However, such
approximations fail to describe the strong-field regime as the black holes merge
\cite{1310.1528}. Numerical relativity is currently the only way to model the
merger event and provide full inspiral-merger-ringdown waveforms. Comparison
between these numerically informed waveforms and detected signals allows GW
observatories to extract physical properties about compact objects \cite{1602.03840, 2009.01190} and measure possible deviations from general relativity \cite{1501.07274, 1602.03841, 1801.03587, 1801.03208}.

Since the first stable BBH simulations in 2005 \cite{gr-qc/0507014,
gr-qc/0511103, gr-qc/0511048}, numerical relativists have enlarged the parameter
space of simulations to include higher spins and mass ratios \cite{Aylott1_skip,
Ajith1_skip, 1304.6077, 1307.5307, 1605.03204, 1703.03423, 1901.02553,
1901.07038, 1904.04831}. Nowadays, for example, the SXS Collaboration has
published BBH simulations of dimensionless spin magnitude up to 0.998 in its
catalog \cite{1904.04831}.  Though all the BBH events published so far by LIGO
and Virgo have effective spin smaller than 0.9 (within 90\% credible intervals)
\cite{1811.12907, 2010.14527}, spin is not constrained to lie in this range and
there is evidence of nearly extremal-spin BHs in X-ray binaries \cite{1106.3690,
1204.5854, 1308.4760, 0902.2840, astro-ph/0606076}. Thus, we must have models of
high-spin BBH GW events so they can be searched for in the detector data.

There are several interesting phenomena that can occur for a high-spin BBH and
not for a non-spinning one. Two examples are the hangup effect \cite{1801.08162}
and the flip-flop effect \cite{1410.3830}. The hangup effect describes how the
merger is delayed or accelerated by spin-orbit coupling compared to a
non-spinning BBH merger, while the flip-flop effect may reverse the spin
direction of a progenitor BH by spin-spin coupling. Both effects can be
exaggerated by near-extremal spins, and can only be further understood by
simulations. Furthermore, high-spin simulations are needed for filling out the
parameter space for surrogate models \cite{1502.07758, 1701.00550, 1705.07089,
1812.07865} and effective-one-body models \cite{gr-qc/9811091, 0711.2628,
1307.6232, 1611.03703}. A recent development among surrogate models is NRSur7dq4
(together with NRSur7dq4Remnant), which is trained with numerical simulations of
spin up to 0.8 \cite{1905.09300}. An instance of effective-one-body calibrations
using simulations of spin up to 0.98 can be found in Ref.~\cite{1311.2544}.

Unfortunately, high-spin simulations are very computationally expensive, and
rapidly increase in cost as the spins are increased. For example, a 25-orbit
spin-0.9 BBH simulation takes weeks to complete, while a 25-orbit spin-0.99
simulation takes months to complete. Thus, a more efficient method of performing
the simulations is highly desirable. Our objective in this paper is to
develop faster high-spin BBH simulations by changing gauge conditions.

All simulations in this paper were done using the spectral Einstein code (SpEC)
\cite{spec_skip}. SpEC uses the first-order generalized harmonic evolution
system to simulate the spacetime \cite{gr-qc/0512093}. Before evolving spacetime
quantities, SpEC solves for the initial data for BBH evolution using the
extended conformal thin-sandwich formalism \cite{gr-qc/9810051, gr-qc/0207095}.
SpEC chooses the free data to be given by a Gaussian-weighted combination of two
single BH analytic solutions \cite{0805.4192}. A traditional choice for the
single BH analytic solution is Kerr-Schild (KS), while more recently
harmonic-Kerr has been used successfully \cite{1808.08228}. Simulations starting
with harmonic-Kerr initial data are up to 30\% faster than ones starting with
KS. However, the harmonic-Kerr initial data can only be solved for spins smaller
than 0.7 \cite{1808.08228}. Reference~\cite{2102.06618} extends harmonic-Kerr initial data to spin-0.9 BBH simulations by using a modified version of harmonic-Kerr, but the overall computational efficiency is not greatly improved compared to simulations using the KS initial data.

In this paper we develop and use several gauge conditions both in the initial
data and in the evolution. We compare their stability, efficiency, and
gravitational waveform output with the goal of making high-spin (dimensionless
spin $\chi\ge0.9$) simulations cheaper. Our most successful choice of initial
data reduces the cost of $\chi=0.99$ aligned-spin simulations by nearly a factor
of 2.

The rest of this paper is organized as follows: in Sec.~\ref{sec:tech},
we describe the numerical methods that are crucial in SpEC BBH simulations and
fundamental in the following discussion of this paper. In Sec.~\ref{sec:config},
we introduce \textit{spherical Kerr-Schild} as a spherical version of KS and
\textit{wide Kerr-Schild} as a modification of spherical Kerr-Schild that
increases the coordinate separation between the inner and outer horizons. We
will also briefly discuss how we delay the transition from the initial data
gauge to the evolution gauge. In Sec.~\ref{sec:results}, we implement these new
configurations in single BH and BBH simulations, and analyze their effect on
computational cost, constraint violations, waveforms, resolution, and apparent
horizons. We finally summarize the results and consider future developments in
Sec.~\ref{sec:conclusion}.

Here are some conventions used in this paper. (1) Unless specified,
\textit{spin} refers to the dimensionless spin $\chi$. Dimensionful spin refers
to spin angular momentum per unit mass and is labeled by $a$. (2) We use
geometric units, i.e.~$G=c=1$. All dimensionful quantities in this paper are
then equipped with units that are an integer power of $M$, the total ADM mass of
a system. For example, time and distance have units of $M$. (3) We use letters
at the beginning of the Latin alphabet ($a,b,c,\dots$) as spacetime indices, and
later letters ($i,j,k,\dots$) as spatial indices. (4) We reserve symbols
$g_{ab}$, $\gamma_{ij}$, $\alpha$, and $\beta^i$ for spacetime metric, spatial
metric, lapse, and shift.

\section{Numerical techniques} \label{sec:tech}

In this section, we provide an overview of some of the numerical methods SpEC
uses. We start by briefly describing the extended conformal thin-sandwich
formalism and SpEC's choice of free data in Sec.~\ref{sec:initialdata}. Next, we
discuss the first-order generalized harmonic system in Sec.~\ref{sec:evolution}.
Finally, in Sec.~\ref{sec:domain}, we briefly describe the configuration of the
computational domain in SpEC.

\subsection{Binary-black-hole initial data} \label{sec:initialdata}

We adopt the standard 3+1 form of the spacetime metric $g_{ab}$,
\begin{align}
  ds^2=-\alpha^2\,dt^2+
  \gamma_{ij}\left(\beta^i\,dt+dx^i\right)\left(\beta^j\,dt+dx^j\right),
\end{align}
where $\alpha$ is the lapse, $\beta^i$ the shift, and $\gamma_{ij}$ the spatial
metric. In vacuum, the spacetime metric $g_{ab}$ and its time derivative
$\partial_tg_{ab}$ must satisfy the Hamiltonian and momentum constraints,
\begin{align}
  R+K^2-K_{ij}K^{ij} &= 0, \\
  D_j(K^{ij}-\gamma^{ij} K) &= 0,
\end{align}
where $R$ is the spatial Ricci scalar, $D_i$ the spatial covariant derivative,
$K_{ij}$ the extrinsic curvature, and $K=K^i{}_i$ the trace of the extrinsic
curvature.

The spatial metric and extrinsic curvature are split using a conformal
decomposition as
\begin{align}
  \gamma_{ij} &= \psi^4\bar{\gamma}_{ij}, \label{eqn:g} \\
  K_{ij} &= A_{ij}+\frac{1}{3}\gamma_{ij}K, \label{eqn:K}
\end{align}
where $\psi$ is the conformal factor, $\bar{\gamma}_{ij}$ the conformal metric,
and $A_{ij}$ the traceless part of $K_{ij}$. $A_{ij}$ is further decomposed as
\begin{align}
  A_{ij} &= \psi^{-2} \bar{A}_{ij}, \\
  \bar{A}^{ij} &= \frac{\psi^6}{2\alpha}\left( (\bar{L}\beta)^{ij} -
                 \bar{u}^{ij} \right),
\end{align}
where $\bar{u}_{ij} \equiv \partial_t \bar{\gamma}_{ij}$ (note that
$\bar{\gamma}^{ij}\bar{u}_{ij} = 0$ to uniquely fix $\bar{u}_{ij}$
\cite{Baumgarte1_skip}), and the vector gradient part $(\bar{L}\beta)^{ij}$ is
defined as
\begin{align}
  (\bar{L}\beta)^{ij} \equiv \bar{D}^i \beta^j + \bar{D}^j \beta^i - \frac{2}{3}
  \bar{\gamma}^{ij} \bar{D}_k \beta^k,
\end{align}
with $\bar{D}_j$ the covariant derivative associated with $\bar{\gamma}_{ij}$.

In the extended conformal thin-sandwich formalism, $\bar{\gamma}_{ij}$, $\bar{u}_{ij}$, $K$, and
$\partial_t K$ are freely specifiable. The elliptic solver in SpEC
\cite{gr-qc/0202096} computes $\psi$, $\alpha$, and $\beta^i$ by solving
\begin{align}
  &\bar{D}_j \left(\frac{\psi^6}{2\alpha}(\bar{L}\beta)^{ij} \right) - \bar{D}_j
    \left(\frac{\psi^6}{2\alpha}\bar{u}^{ij} \right) - \frac{2}{3}\psi^6
    \bar{D}^i K = 0, \\
  &\bar{D}^2\psi - \frac{1}{8}\psi\bar{R} - \frac{1}{12}\psi^5K^2 +
    \frac{1}{8}\psi^{-7} \bar{A}_{ij}\bar{A}^{ij} = 0, \\
  &\bar{D}^2(\alpha\psi) -
    \alpha\psi\left(\frac{7}{8}\psi^{-8}\bar{A}_{ij}\bar{A}^{ij} +
    \frac{5}{12}\psi^4K^2 +\frac{1}{8}\bar{R} \right) \nonumber \\
  &\hspace{0.47in} + \psi^5(\partial_t K - \beta^k\partial_k K) = 0,
\end{align}
where $\bar{R}$ is the conformal Ricci scalar. Equations~\eqref{eqn:g} and
\eqref{eqn:K} are then used to compute $\gamma_{ij}$ and $K_{ij}$.

The free variables $\bar{u}_{ij}$ and $\partial_t K$ are typically set to 0 to
construct quasi-equilibrium initial data. SpEC sets the remaining free variables
$\bar{\gamma}_{ij}$ and $K$ using a superposition of two single BH spacetimes
blended together by a Gaussian weight function \cite{0805.4192}. We define
$\gamma^\rho_{ij}$ and $K^\rho$ to refer to these quantities for
a boosted spinning BH, where $\rho = A, B$ labels each BH. $\bar{\gamma}_{ij}$ and $K$ are chosen to be \cite{0805.4192}
\begin{align}
\bar{\gamma}_{ij} &\equiv \eta_{ij} + \sum_\rho e^{-r^2_\rho/w^2_\rho}(\gamma^\rho_{ij}-\eta_{ij}), \\
K &\equiv \sum_\rho e^{-r^2_\rho/w^2_\rho}K^\rho,
\end{align}
where $\eta_{ij}$ is the 3D flat metric, $r_\rho$ is the Euclidean distance
from BH $\rho$, and $w_\rho$ controls the falloff of BH $\rho$'s
contribution. We use $w_\rho$ equal to 3/5 of the Euclidean distance
between the BBH's L1 Lagrange point (Euclidean center-of-mass) and BH $\rho$'s
center. This choice of $w_\rho$ is wider than a BH's size but still relatively
far from the companion BH.

The most common choice for the free data at each BH is Kerr-Schild (KS), though
using harmonic-Kerr has much promise for low-spin binaries \cite{1808.08228}. In
this paper, we use a Kerr-Schild-like gauge where the horizon is spherical at
each black hole to set the free data. We will discuss the KS and KS-like gauges
in Sec.~\ref{sec:config}.

\subsection{Generalized harmonic evolution system} \label{sec:evolution}

SpEC evolves the initial data using the first-order generalized harmonic (GH)
system \cite{gr-qc/0512093}. (See Refs.~\cite{Friedrich1_skip, gr-qc/0110013,
  gr-qc/0407110} for more details on the GH systems.) The coordinates $x^a$
(referred to as generalized harmonic coordinates) satisfy the inhomogeneous wave
equation,
\begin{align}
  H^a = \nabla_b \nabla^b x^a = -\Gamma^a,
\end{align}
where $\Gamma^a\equiv g^{bc}\Gamma^a_{bc}$ is the trace of the Christoffel
symbol, and $\nabla_a$ the $g_{ab}$-compatible covariant derivative. The gauge
source function $H^a=H^a(x^b,g_{cd})$ is any arbitrary function dependent only
on $x^b$ and $g_{cd}$ (but not derivatives of $g_{ab}$). In these coordinates,
the vacuum Einstein equations can be cast into a manifestly hyperbolic form,
\begin{align}
  g^{cd}\partial_c\partial_d g_{ab} = &-2\nabla_{(a} H_{b)} \nonumber \\
                                      &+2g^{cd}g^{ef}(\partial_e
                                        g_{ca}\partial_f g_{db}
                                        -\Gamma_{ace}\Gamma_{bdf}),
                                        \label{eqn:2ndGH}
\end{align}
where $\Gamma_{abc} = g_{ad}\Gamma^d_{bc}$. After expanding
Eq.~\eqref{eqn:2ndGH} into a first-order representation (done analogously to
expanding the covariant scalar field system \cite{gr-qc/0305027, gr-qc/0407011})
and adding constraint damping terms (see \cite{gr-qc/0402027, gr-qc/0407110,
  gr-qc/0504114, gr-qc/0507014, gr-qc/0512093} for detailed discussions on
constraint damping), we arrive at the GH evolution equations implemented in
SpEC,
\begin{align}
  \partial_{t} g_{a b}&=-\alpha \Pi_{a b}-\gamma_{1} \beta^{i} \Phi_{i a b}
                        +\left(1+\gamma_{1}\right) \beta^{k} \partial_{k} g_{a
                        b}, \label{eqn:GH1} \\
  \partial_{t} \Pi_{a b} &=2 \alpha g^{c d}\left(g^{i j} \Phi_{i c a} \Phi_{j d
                           b}-\Pi_{c a} \Pi_{d b}-g^{e f} \Gamma_{a c e}
                           \Gamma_{b d f}\right) \nonumber \\
                      &-2 \alpha \nabla_{(a} H_{b)}-\frac{1}{2} \alpha n^{c}
                        n^{d} \Pi_{c d} \Pi_{a b}-\alpha n^{c} \Pi_{c i} g^{i j}
                        \Phi_{j a b} \nonumber \\
                      &+\alpha \gamma_{0}\left[2 \delta^{c}{ }_{(a} n_{b)}-(1+\gamma_{3})g_{a
                        b} n^{c}\right]\left(H_{c}+\Gamma_{c}\right) \nonumber  \label{eqn:GH2} \\
                      &-\gamma_{1} \gamma_{2} \beta^{i} \Phi_{i a b} +\beta^{k} \partial_{k} \Pi_{a b} -\alpha g^{k i}
                        \partial_{k} \Phi_{i a b} \nonumber \\
                      &+\gamma_{1} \gamma_{2}
                        \beta^{k} \partial_{k} g_{a b}, \\
  \partial_{t} \Phi_{i a b} &=\frac{1}{2} \alpha n^{c} n^{d} \Phi_{i c d} \Pi_{a
                              b}+\alpha g^{j k} n^{c} \Phi_{i j c} \Phi_{k a
                              b}-\alpha \gamma_{2} \Phi_{i a b} \nonumber\\
                      &+\beta^{k} \partial_{k} \Phi_{i a b}-\alpha \partial_{i}
                        \Pi_{a b}+\alpha \gamma_{2} \partial_{i} g_{a b},  \label{eqn:GH3} 
\end{align}
where $g_{ab}$, $\Phi_{iab} \equiv \partial_i g_{ab}$,
$\Pi_{ab} \equiv -n^c\partial_c g_{ab}$ are the three dynamical fields being
evolved, $\gamma_0$, $\gamma_1$, $\gamma_2$ and $\gamma_3$ the constraint
damping parameters, and $n^a$ the future-pointing unit normal to constant-$t$
spatial hypersurfaces. See the Appendix for values of $\gamma_0$, $\gamma_1$, $\gamma_2$ and $\gamma_3$ used in the simulations of this paper.

The simplest choice of gauge source function is the harmonic gauge, where
$H^a=0$. The harmonic gauge dates back to Einstein's work \cite{Renn1_skip} and
has been an important tool in many aspects of analytical general relativity
\cite{Donder1_skip, Fock1_skip, Fischerr1_skip}. Unfortunately, using a harmonic
gauge condition in simulations of BBH mergers leads to explosive growth of
$\gamma=\det(\gamma_{ij})$ near the apparent horizons as two BHs merge
\cite{0909.3557}. To suppress such growth, SpEC adopts the damped wave gauge or
damped harmonic (DH) gauge $H^a = H^a_{\text{DH}}$ \cite{0909.3557},
\begin{align}
  &H^a_{\text{DH}} = \mu_L n^a \ln\left(\frac{\sqrt{\gamma}}{\alpha}\right) - \mu_S \frac{\beta^i}{\alpha} \gamma^a_{\ i}, \\
  &\mu_L = \mu_S = e^{-(\ln10^{15})r^2/\sigma^2} \left[ \ln\left(\frac{\sqrt{\gamma}}{\alpha}\right)\right]^2,
\end{align}
where $r$ is the Euclidean radius, and $\sigma=100M$. Note that the DH gauge
reduces to the harmonic gauge to machine precision for $r\ge\sigma$.

SpEC uses the following gauge transition from the initial gauge
$H_{\text{init}}^a$ to the DH gauge $H^a_{\text{DH}}$ during evolution:
\begin{align}
  H^a = F(t)H^a_{\mathrm{init}}+\left[ 1-F(t) \right]H^a_{\mathrm{DH}},
\end{align}
where
\begin{align}
  F(t) =
  \begin{dcases}
    \exp\left[-\left(\frac{t-t_0}{w}\right)^4\right], & t\ge t_0 \\
    1, & t<t_0.
  \end{dcases} \label{eqn:time_transition}
\end{align}
Here $t_0$ is the start time and $w$ is the temporal width of the
transition. Unless stated otherwise, we choose $t_0=0M$ and $w=50M$.
Note that this transition function is not finely tuned. It
is chosen because it decays to 0 rapidly for large $t$ and has a
continuous third derivative at $t=t_0$.

\subsection{Computational domain} \label{sec:domain}

SpEC adopts a dual-frame configuration for BBH simulations
\cite{gr-qc/0607056}. In the \textit{inertial frame}, the two BHs are orbiting
each other and deform as they merge. In contrast, in the \textit{grid frame}, the
BHs are at fixed coordinate locations and are kept approximately spherical. The
two frames are related by a time-dependent analytic map determined by feedback
control systems in SpEC \cite{1211.6079}.

SpEC's domain decomposition is described in the Appendix of
Ref.~\cite{1206.3015}, while the adaptive mesh refinement (AMR) algorithm is
described in Refs.~\cite{1010.2777, 1405.3693}. SpEC uses spherical shells
around each BH. The number of spherical harmonic modes ($\ell$) used in the
shells around BHs is a direct proxy for how the shape of each BH affects the
computational cost of a simulation. High-spin BBH simulations use $\ell\ge40$,
which results in not only many grid points, but also close spacing between grid
points. A simulation with more grid points requires more computation per time
step, while a closer spacing between grid points requires a smaller time step in
order to maintain stability. Both factors slow down the overall simulation. With
this in mind, we seek to reduce the angular resolution needed in high-spin BBH
simulations, anticipating faster simulations.

SpEC uses excision to avoid the physical singularities inside BHs. Specifically,
the region within an inner boundary for each BH is excluded from the
computational domain. This boundary is called an \textit{excision surface} or
\textit{excision boundary} and lies slightly inside the apparent horizon (AH) of
each BH \cite{gr-qc/0407110, 1211.6079}. Causality prohibits any physical
content in the interior region from propagating out.
The excision surface has to be
placed in a trapped region between the inner and outer horizons for each
BH. Since the distance between the inner and outer horizons decreases as spin
increases, the placement of the excision boundary becomes increasingly difficult
as the spin increases. As a result, smaller time step sizes are necessary to
track the apparent horizons and to keep the excision boundary inside the narrow
trapped region. Thus, a gauge where horizons remain spherical for any spin
should not only decrease the resolution used by AMR but also reduce the workload
in tracking the apparent horizons and controlling the excision boundaries. At the outer boundary, suitable constraint-preserving boundary conditions \cite{gr-qc/0512093} are imposed.

\section{New initial data and gauge conditions} \label{sec:config}

We describe three modifications to SpEC's current configuration that will be
explored in this paper. The major modification is to introduce a
Kerr-Schild-like gauge in the free initial data, where the horizons are
spherical for any spin. We will refer to this choice as \textit{spherical
Kerr-Schild}. The other two modifications are variants of the spherical KS
initial data. One variant increases the coordinate distance between the inner
and outer horizons in spherical Kerr-Schild to construct what we refer to as the
\textit{wide Kerr-Schild} gauge. The other variant keeps the spherical
Kerr-Schild data, but delays the transition to damped harmonic gauge during the
evolution.

\subsection{Spherical Kerr-Schild} \label{sec:sphks_coord}

The Kerr metric in KS coordinates $\{t,x,y,z\}$, with spin pointing along the
$z$-axis,\footnote{For spin not along the $z$-axis adding a 3D rotation suffices
  to determine the metric.} mass $M$, and angular momentum $aM=\chi M^2$ is
\begin{align}
  g_{ab} = \eta_{ab} + 2H l_a l_b,
\end{align}
where $\eta_{ab}$ is the Minkowski metric,
\begin{align}
  H &= \frac{Mr^3}{r^4+a^2 z^2}, \\
  l_a &= \left( 1, \frac{rx+ay}{r^2+a^2}, \frac{ry-ax}{r^2+a^2}, \frac{z}{r} \right),
\end{align}
and $r$ implicitly given by
\begin{align}
  \frac{x^2+y^2}{r^2+a^2}+\frac{z^2}{r^2}=1 \label{eqn:spheroid}
\end{align}
is the radial coordinate in Boyer-Lindquist coordinates. Since we are only
interested in astrophysical BHs, we restrict ourselves to $\chi<1$.  Then for
any nonzero spin, there are inner and outer horizons
$r_\pm=(1\pm\sqrt{1-\chi^2})M$. In the region $r_-<r<r_+$, any object must
travel radially inward, while outside this region ($r<r_-$ or $r>r_+$, excluding
horizons), an object can travel both radially inward and outward.  For high
spins, the horizons $r_{\pm}$ become closer together and increasingly
non-spherical in the KS coordinate system, requiring greater angular resolution
to simulate the BHs. This suggests that we might mitigate the required
resolution increase by using coordinates in which the horizons are spherical, as
we now describe.

\begin{figure}[t]
	\includegraphics[width=\linewidth]{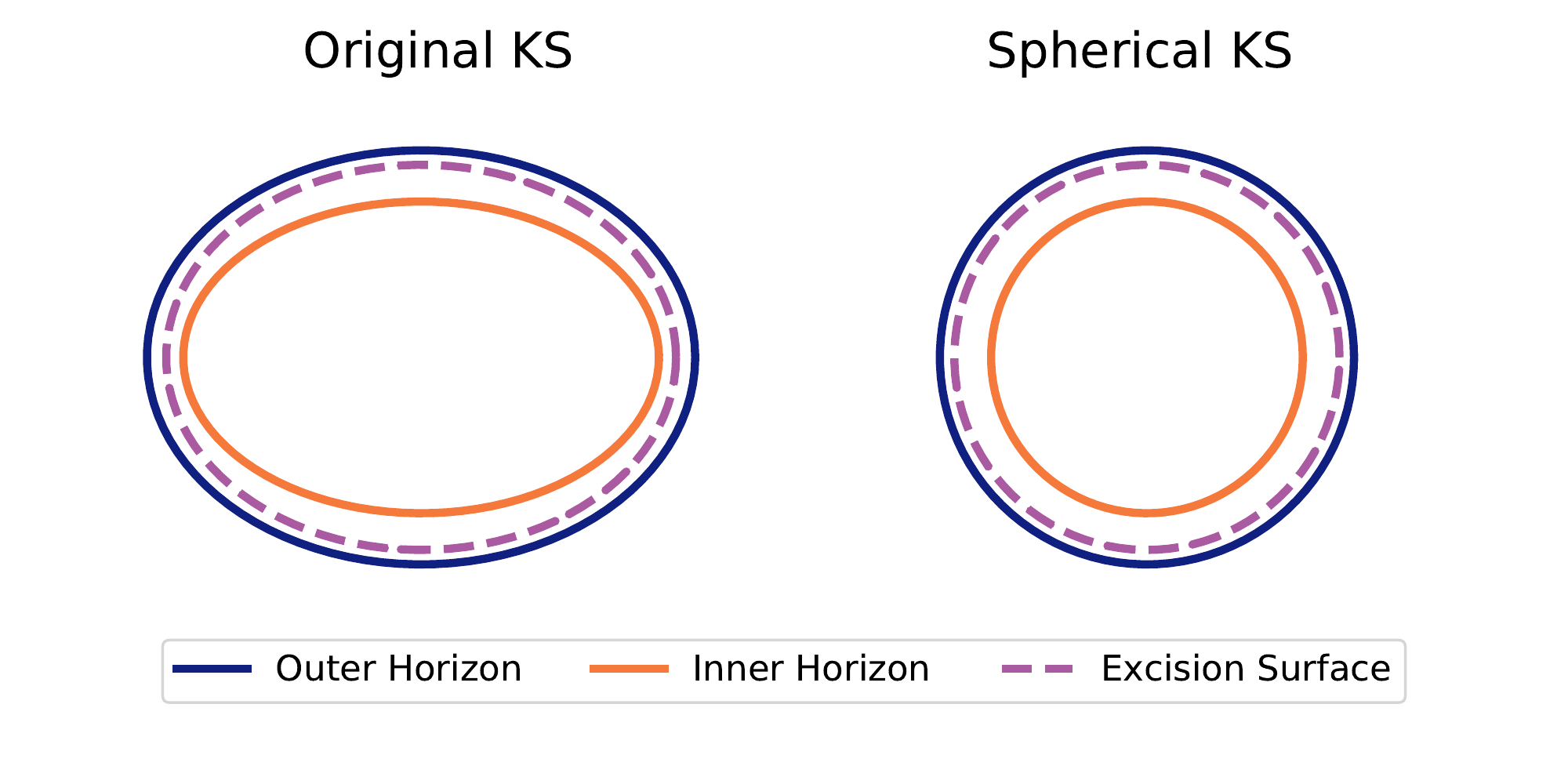}
	\caption{The $xz$-plane for a single BH with spin 0.9 along the $z$-axis in KS
		coordinates (left) and spherical KS coordinates (right). The left diagram
		shows that the excision surface and both horizons are spheroids.  The right
		one shows that in spherical KS coordinates the excision surface and horizons
		are spheres.}
	\label{fig:coordtrans}
\end{figure}

We denote the spherical KS coordinates by $\{ t,\bar{x},\bar{y},\bar{z}
\}$. These are related to the KS coordinates $\{t,x,y,z\}$ by
\begin{align}
	\frac{\bar{x}}{r} &= \frac{x}{\sqrt{r^2+a^2}}, \label{eqn:coordtransf1}\\
	\frac{\bar{y}}{r} &= \frac{y}{\sqrt{r^2+a^2}}, \label{eqn:coordtransf2}\\
	\bar{z} &= z. \label{eqn:coordtransf3}
\end{align} 
Equation~\eqref{eqn:spheroid} describes an oblate spheroid in $\{x,y,z\}$ and is
equivalent to a sphere in $\{\bar{x},\bar{y},\bar{z}\}$ coordinates. That is,
\begin{align}
  \bar{x}^2+\bar{y}^2+\bar{z}^2=r^2. \label{eqn:sphks_sphere}
\end{align}
The left panel of Fig.~\ref{fig:coordtrans} shows the inner and outer horizons
in KS as solid lines, and a sample excision surface as a dashed line.
The right panel of Fig.~\ref{fig:coordtrans} shows the inner and outer horizons,
and excision surface but in spherical KS coordinates instead. We will abbreviate
spherical KS as SphKS hereinafter.

\subsection{Wide Kerr-Schild} \label{sec:wks}

Recall from Sec.~\ref{sec:domain} that as the BHs inspiral, the excision regions
track the BHs. The excision boundary must be inside $r_{+}$ but outside $r_{-}$
so that all information leaves the computational domain and no boundary
condition must be applied. This becomes more difficult as the spin increases,
partly because the space between $r_{+}$ and $r_{-}$ decreases. We attempt to
reduce the work of the control system by expanding the region between the
horizons by performing a radial transformation. Note that the idea of expanding the region between horizons in the initial data is not new. For example, Ref.~\cite{1706.01980} applies a fisheye radial transformation to the quasi-isotropic coordinates to expand the horizon size. We here apply a different radial transformation to the SphKS coordinates. We
refer to this gauge as \textit{wide Kerr-Schild} (WKS). We continue using the
notation in Sec.~\ref{sec:sphks_coord} and denote the coordinates of WKS as
\{$t,\tilde{x},\tilde{y},\tilde{z}$\}. We introduce a new variable $\tilde{r}$
that is related to $r$ and choose the coordinate transformation between WKS and
SphKS as
\begin{align}
  \frac{\tilde{x}}{\tilde{r}} &= \frac{\bar{x}}{r}, \\
  \frac{\tilde{y}}{\tilde{r}} &= \frac{\bar{y}}{r}, \\
  \frac{\tilde{z}}{\tilde{r}} &= \frac{\bar{z}}{r}.
\end{align}
With this convention, Eqs.~\eqref{eqn:spheroid} and \eqref{eqn:sphks_sphere} are
equivalent to
\begin{align}
  \tilde{x}^2+\tilde{y}^2+\tilde{z}^2=\tilde{r}^2,
\end{align}
i.e.~a sphere of radius $\tilde{r}$ in WKS.

Starting with SphKS, we want a radial transformation $r \rightarrow \tilde{r}$
that keeps $r_{+}$ fixed but shrinks $r_{-}$ radially inward by some factor $b$,
i.e.,
\begin{align}
	\tilde{r}(r_+) &= r_+, \\
	\tilde{r}(r_-) &= br_-.
\end{align}
We can achieve these relations with a quadratic,
\begin{align}
  \tilde{r}(r)=\frac{1-b}{r_{+}-r_{-}}r^2 + \frac{br_{+}-r_{-}}{r_+-r_{-}}r,
  \label{eqn:quadratic}
\end{align}
with $r_-/r_+\le b\le1$ to ensure monotonicity, e.g., the lower bound of $b$
is $\sim0.75$ for a spin-0.99 BH. This $b$ is a \textit{squeezing parameter}
that controls how far $r_-$ is pushed inwards after the transformation.

\begin{figure}[t]
  \centering \includegraphics[width=\linewidth]{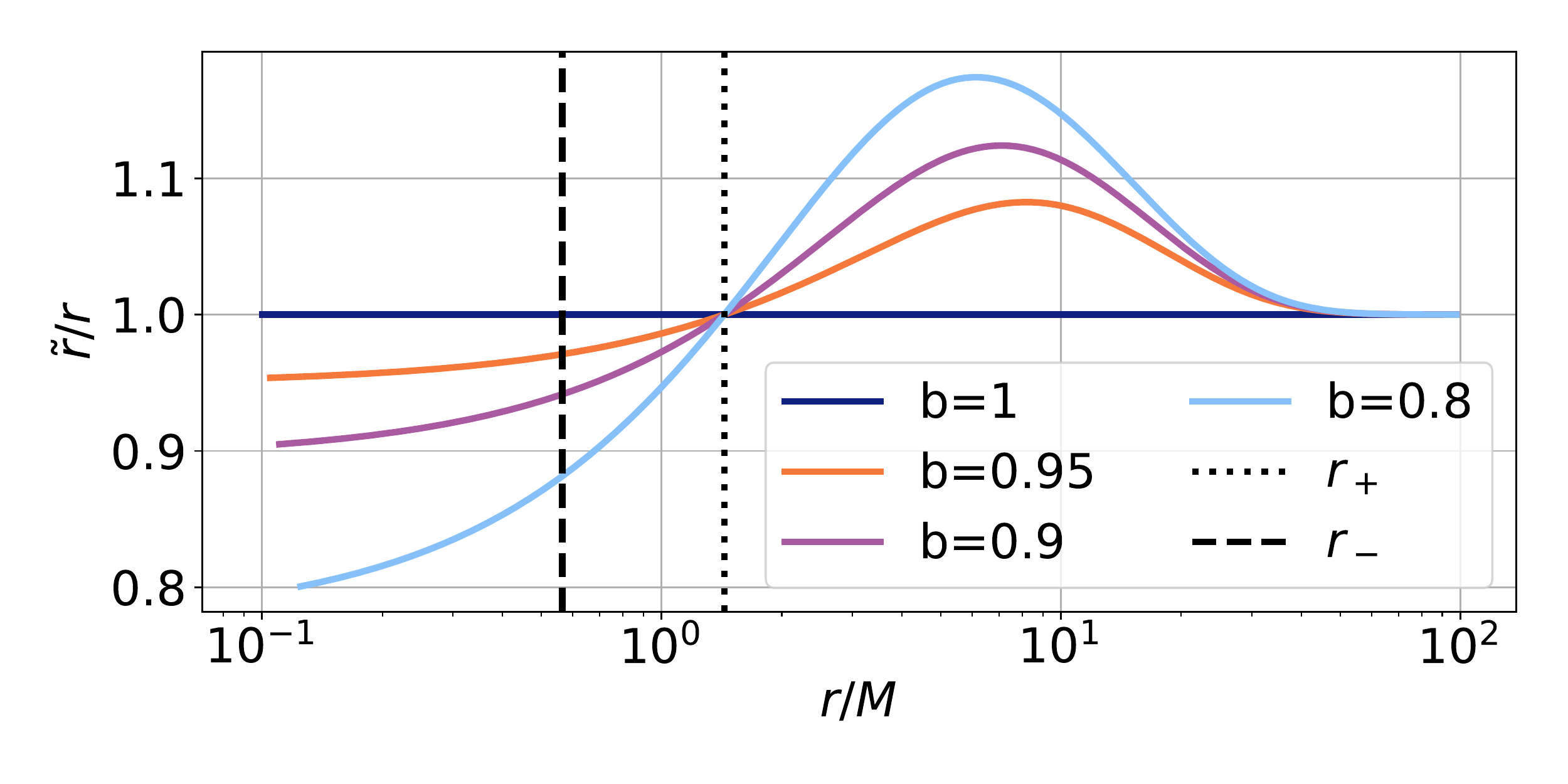}
  \caption{The ratio $\tilde{r}/r$ as a function of $r$ for various squeezing
	parameters $b$ for a spin-0.9 BH with $C=3M$ and $\lambda=8M$. The horizontal
	axis ($r/M$) is on a logarithmic scale to show both near-field and far-field
	behaviors. Since for $b=1$ (no squeezing) $\tilde{r}=r$, the curve stays at 1.
	All curves tend to 1 far from the BH to preserve asymptotic flatness, while
	becoming smaller than 1 inside the outer horizon. The reader can confirm from
	the graph that the transformation in Eq.~\eqref{eqn:wks quadratic} keeps $r_+$
	fixed.}
  \label{fig:wks}
\end{figure}

Unfortunately, this quadratic has the pathology that the metric would no longer
be asymptotically flat, so we use the quadratic only near the BH, smoothly
transitioning to $\tilde{r}=r$ far away from the BH. Specifically,
\begin{align}
	r(\tilde{r}) =
  \frac{\sqrt{B^2+4A\tilde{r}}-B-2A\tilde{r}}{2A}\frac{1}{1+e^{(\tilde{r}-C)/\lambda}}
  + \tilde{r},
  \label{eqn:wks quadratic}
\end{align}
where
\begin{align}
	A &= \frac{1-b}{r_+-r_-}, \\
	B &= \frac{br_+-r_-}{r_+-r_-} = 1-Ar_+.
\end{align}
We write the relation as a function $r(\tilde{r})$ instead of $\tilde{r}(r)$ for
easier implementation in SpEC. $C$ and $\lambda$ are parameters of the sigmoid
function, controlling the center and width of transition. Theoretically, we
could choose $C$ and $\lambda$ to satisfy $e^{(r_+-C)/\lambda}\sim10^{-15}$ so
that Eq.~\eqref{eqn:quadratic} holds exactly inside $r_+$ within numerical
precision, but such combinations always result in large Jacobians. A quantity
with large derivatives (and second derivatives) needs sufficiently high
resolution to be resolved, which increases computational cost. In practice,
because the starting point of WKS is broadening the region between $r_+$ and
$r_-$ nonlinearly, we simply choose $C=3M$ and $\lambda=8M$.  This combination
of $C$ and $\lambda$ maintains stable simulations without increasing
computational cost too much. Figure~\ref{fig:wks} shows $\tilde{r}/r$ versus $r$
for these $C$ and $\lambda$.

\subsection{Delayed evolution gauge transition} \label{sec:delay}

When studying evolutions using SphKS initial data we noticed that the apparent
horizons become spheroids around $t=50M$ when the gauge has mostly transitioned
to damped harmonic. With the change to a spheroidal horizon we observe the
expected increase in the required angular resolution. Since the damped harmonic
(DH) condition is generally only necessary during merger, we perform simulations
where we delay the transition from $H^a_{\mathrm{init}}$ to $H^a_{\mathrm{DH}}$
in an attempt to extend how long the horizons remain (nearly) spherical. We
change both $t_0$ and $w$ in Eq.~\eqref{eqn:time_transition} and report our
results in Sec.~\ref{sec:spin0.9bbh_delay}.

\section{Results} \label{sec:results}

In this section, we test the new configurations described in
Sec.~\ref{sec:config} by evolving multiple single BH and BBH systems. In
particular, we will compare the constraint violations, computational efficiency,
AH shapes, total number of grid points, and waveforms from different
simulations.

\subsection{Single spherical Kerr-Schild black hole} \label{sec:sbh}

We evolve a spin-0.99 BH to time $4000M$ in both KS and SphKS coordinates. We
keep the domain decomposition the same by using the transformation
Eqs.~(\ref{eqn:coordtransf1}$-$\ref{eqn:coordtransf3}) from the SphKS domain,
such that any sphere is mapped to a spheroid matching the given spin. AMR is
disabled, ensuring the domain decomposition and resolution are unchanged
throughout the simulations. We fix the radial resolution in both coordinates but
vary the angular resolutions, represented by $\ell$. The number of angular grid
points is then $2(\ell+1)^2$. We investigate four single BH simulations: three in KS with $\ell=22,26,30$ and one in SphKS with $\ell=22$.

\begin{figure}[t]
  \centering \includegraphics[width=\linewidth]{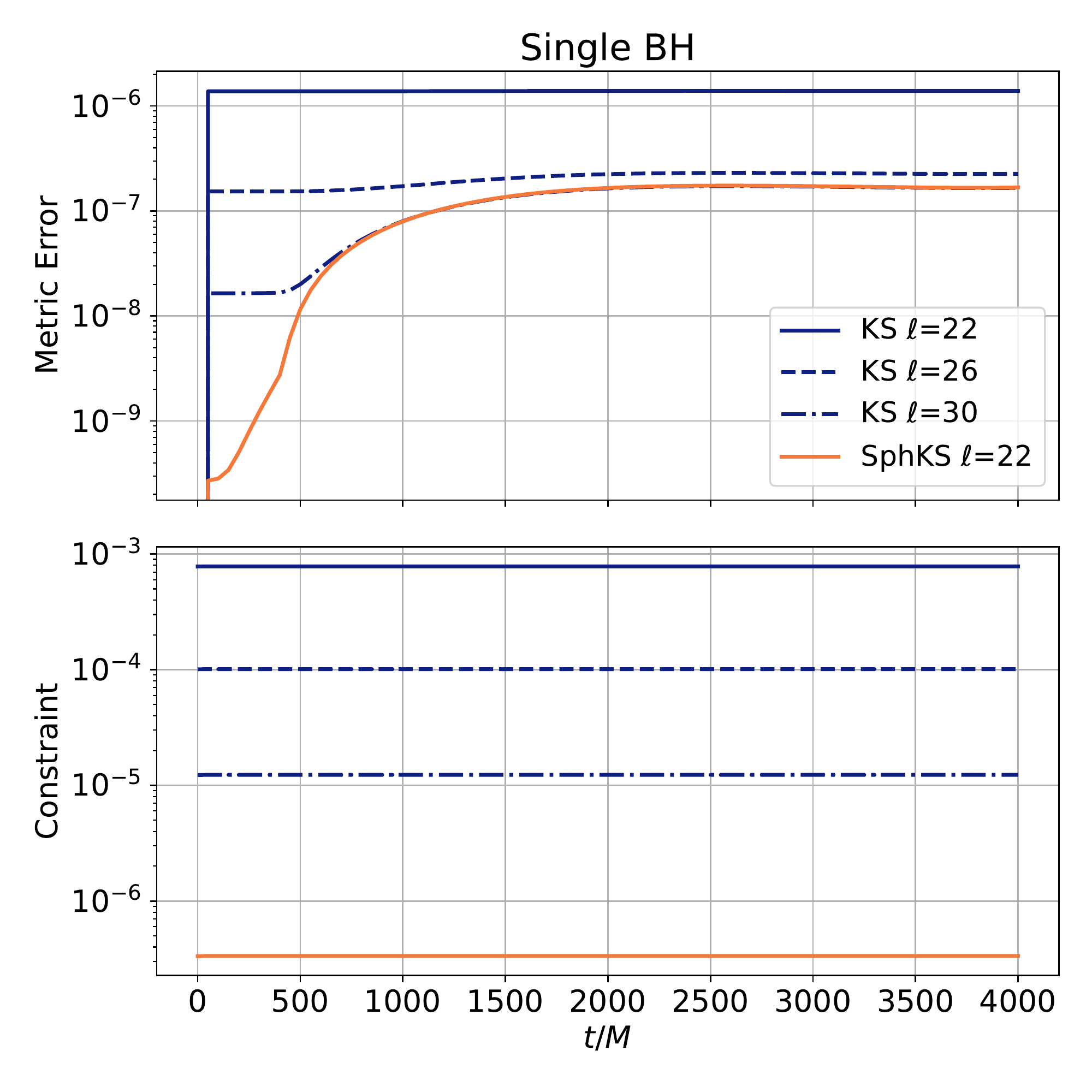}
  \caption{Metric errors and constraint energy of four single BH simulations.
    The top panel plots the $L_2$-norm of the error of the spacetime metric,
    i.e.~the $L_2$-norm of $g_{ab} - g_{ab}^{\mathrm{analytic}}$. The bottom
    panel plots the $L_2$-norm of the constraint energy. The blue lines
    represent the simulations using KS initial data, with $\ell=22,26,30$ in the
    solid, dashed, and dash-dotted line styles, while the orange solid lines are
    results from the SphKS $\ell=22$ simulation. As $\ell$ increases, the KS
    simulations achieve lower metric errors and constraint violations. The SphKS
    $\ell=22$ simulation has lower metric errors and constraint violations than
    the KS $\ell=22$ case by factors of 10 and $10^3$. Note that the $10^{-7}$
    error floor of the metric errors at late times is caused by the outer
    boundary condition.}
  \label{fig:sbh}
\end{figure}

In the top panel of Fig.~\ref{fig:sbh}, we show the $L_2$-norm of the metric
errors $g_{ab}-g_{ab}^{\mathrm{analytic}}$ for all four simulations. The blue lines
represent the evolution using KS coordinates, with $\ell=22,26,30$ corresponding
to the solid, dashed and dash-dotted styles. The metric error decreases as the
angular resolution increases, especially at early time (before $t<500M$). After
$t\sim1000M$, the metric error approaches a $10^{-7}$ error floor. This error
floor appears because the numerical error gets reflected at the outer boundary
and amplified when traveling inward \cite{0704.0782}.\footnote{The floor is
decreased when we move the outer boundary farther out.} The orange line is for
the evolution using SphKS coordinates with $\ell=22$.\footnote{Simulations in
SphKS coordinates with higher $\ell$'s yield nearly the same metric errors as
the $\ell=22$ curve, so they are not shown.} It reaches the same error floor at
late time but is at least 10 times smaller than the metric error of the
$\ell=22$ KS simulation (the blue solid curve).

In the bottom panel of Fig.~\ref{fig:sbh}, we show the GH constraint energy $\mathcal{E}_{c}$ for these four simulations. The constraint energy (or
\textit{constraint violation}) used in this paper is defined as
\begin{align}
\mathcal{E}_{c} = \displaystyle \sqrt{\frac{\int \mathcal{C}_{\mathrm{GH}}^2 \sqrt{\gamma}d^3x}{\int \sqrt{\gamma}d^3x}}, \label{eqn:ghce}
\end{align}
where
\begin{align}
	\mathcal{C}_{\mathrm{GH}}^2 = \delta^{a b} &[ \gamma^{i j}(\mathcal{C}_{i a} \mathcal{C}_{j b} + \delta^{c d} \mathcal{C}_{i a c} \mathcal{C}_{j b d}
	+ \gamma^{k l} \delta^{c d} \mathcal{C}_{i k a c} \mathcal{C}_{j l b d}) \nonumber \\
	&+  \mathcal{F}_{a} \mathcal{F}_{b}	+ \mathcal{C}_{a} \mathcal{C}_{b}].
\end{align}
Here, $\delta^{a b}$ is the 4D Kronecker delta and \{$\mathcal{C}_{a}$,
$\mathcal{F}_{a}$, $\mathcal{C}_{i a}$, $\mathcal{C}_{i a b}$, $\mathcal{C}_{i j
a b}$\} are the five constraints used in Ref.~\cite{gr-qc/0512093}. Note that our
definition of $\mathcal{E}_{c}$ is different from the one in
Ref.~\cite{gr-qc/0512093}. Among the simulations using KS coordinates, the
constraint violation is smaller as $\ell$ increases. In contrast to the metric
errors, the constraint violations do not hit an error floor because of the use
of a constraint-preserving boundary condition \cite{0704.0782}. We see that the
SphKS evolution has constraint violation over a factor of $10^3$ smaller than
the evolution in KS coordinates using the same angular resolution ($\ell=22$).
Even compared to the KS $\ell=30$ case, the SphKS $\ell=22$ simulation still has a smaller constraint violation by a factor of 10.

\begin{table*}
	\caption{Parameters for the six BBH simulations studied in
		Secs.~\ref{sec:spin0.9bbh} and \ref{sec:spin0.99bbh}. $\vec{\chi}_{A,B}$ are
		the spin vectors of the progenitor-BHs. $D_0$, $\Omega_0$, $\dot{a}_0$ and $e$ are the initial coordinate separation, initial orbital frequency,
		initial rate of change of separation, and eccentricity. All these six simulations have mass ratio 1.}
	\label{tbl:bbh_param}
	\begin{ruledtabular}
		\begin{tabular}{ccccccc}
			$\vec{\chi}_{A,B}$ & Initial data gauge  & $D_0$ $[M]$ & $\Omega_0$ & $\dot{a}_0$ & $e$ & \texttt{\#} of orbits  \\ 
			\hline \noalign{\vskip 1mm} 
			(0, 0, 0.9) & KS & 15.450 & 1.4095$\times 10^{-2}$ & 5.3578$\times 10^{-4}$ & $\sim0.0003$ & $\sim25$ \\
			& SphKS & 15.450 & 1.42$\times 10^{-2}$ & 4.5284$\times 10^{-4}$ & $\sim0.0005$ & $\sim25$ \\
			(0, 0, 0.95) & KS & 11.580 & 2.0875$\times 10^{-2}$ & 1.0650$\times 10^{-3}$ & $\sim0.0007$ & $\sim14$ \\
			& SphKS & 11.580 & 2.1100$\times 10^{-2}$ & 8.5921$\times 10^{-4}$ & $\sim0.0005$ & $\sim14$ \\
			(0, 0, 0.99) & KS & 11.577 & 2.0384$\times 10^{-2}$ & 1.4799$\times 10^{-3}$ & $\sim0.0005$ & $\sim14$ \\
			& SphKS & 11.577 & 2.0808$\times 10^{-2}$ & 1.1357$\times 10^{-3}$ & $\sim0.0002$ & $\sim14$ \\
		\end{tabular}
	\end{ruledtabular}
\end{table*}

Because the horizons are spherical in the SphKS gauge, spacetime quantities are
constructed and evolved directly in spherical domains. In the KS gauge,
quantities are evolved in spheroidal domains, so a spatial map converting
spheroids to spheres is necessary in the spectral calculation of derivatives.
The Jacobian and Hessian of this spatial map and its inverse can introduce
errors. Thus, in Fig.~\ref{fig:sbh}, we see the single BH simulation in SphKS
provides a more accurate result than in KS.

The KS $\ell=22$ simulation takes 978 CPU hours to reach $t=4000M$, while the
SphKS $\ell=22$ simulation takes 948 CPU hours on the same number of cores.
However, a better comparison is to the $\ell=26$ KS case, which takes 1438 CPU
hours and still has considerably larger errors. It may not be surprising that we
are able to reduce the numerical error of single BH simulations by using
coordinates better adapted to the geometry of the BH, but it is reassuring to
have confirmation.

\subsection{Spin-0.9 binary-black-hole simulations using spherical Kerr-Schild
  initial data} \label{sec:spin0.9bbh}

We evolve three pairs of non-eccentric, non-precessing, equal-mass, equal-spin BBH
systems, corresponding to spin 0.9, 0.95, and 0.99, all along the $z$-axis. Each
pair consists of a run with superposed KS initial data and another run with
superposed SphKS initial data. They both merge after nearly the same number of
orbits and at nearly the same simulation time. The initial orbital frequency
$\Omega_0$ and the initial rate of change of separation $\dot{a}_0$ are tuned
separately for each run, subject to a fixed initial separation $D_0$. We perform
this tuning by eccentricity reduction \cite{1012.1549} to achieve a negligible
eccentricity $(e<0.0007)$. The specific values of these parameters, including
the number of orbits, are provided in Table~\ref{tbl:bbh_param}. Furthermore, we
simulate each BBH run at three resolutions, Lev-1, Lev-2, and Lev-3. For
Lev-$i$, the target truncation error of the AMR algorithm is
$\sim 2\times 4^{-i} \times 10^{-4}$. 

\begin{figure*}[t]
  \centering
  \includegraphics[width=0.49\linewidth]{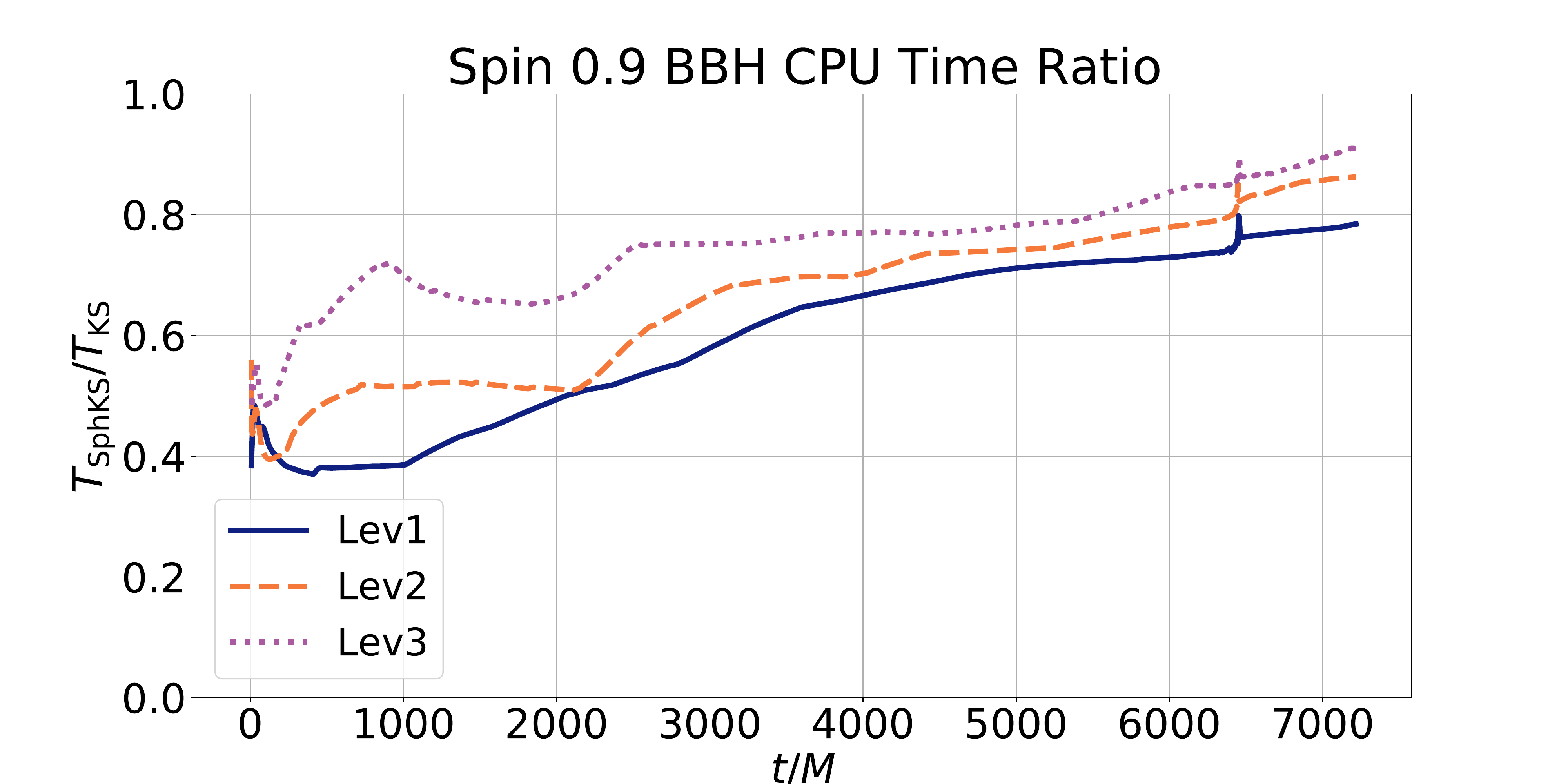}
  \includegraphics[width=0.49\linewidth]{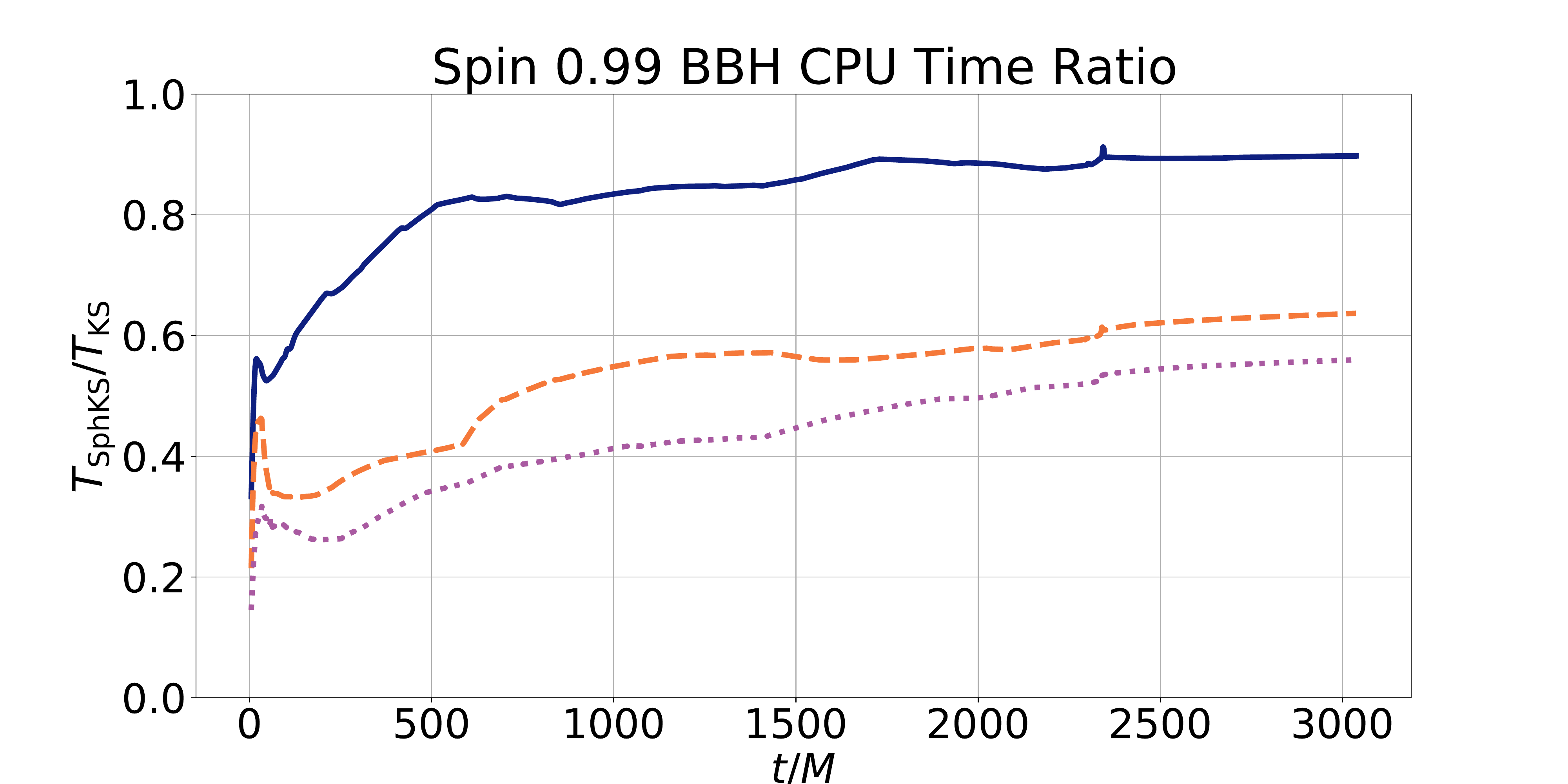}
  \includegraphics[width=0.49\linewidth]{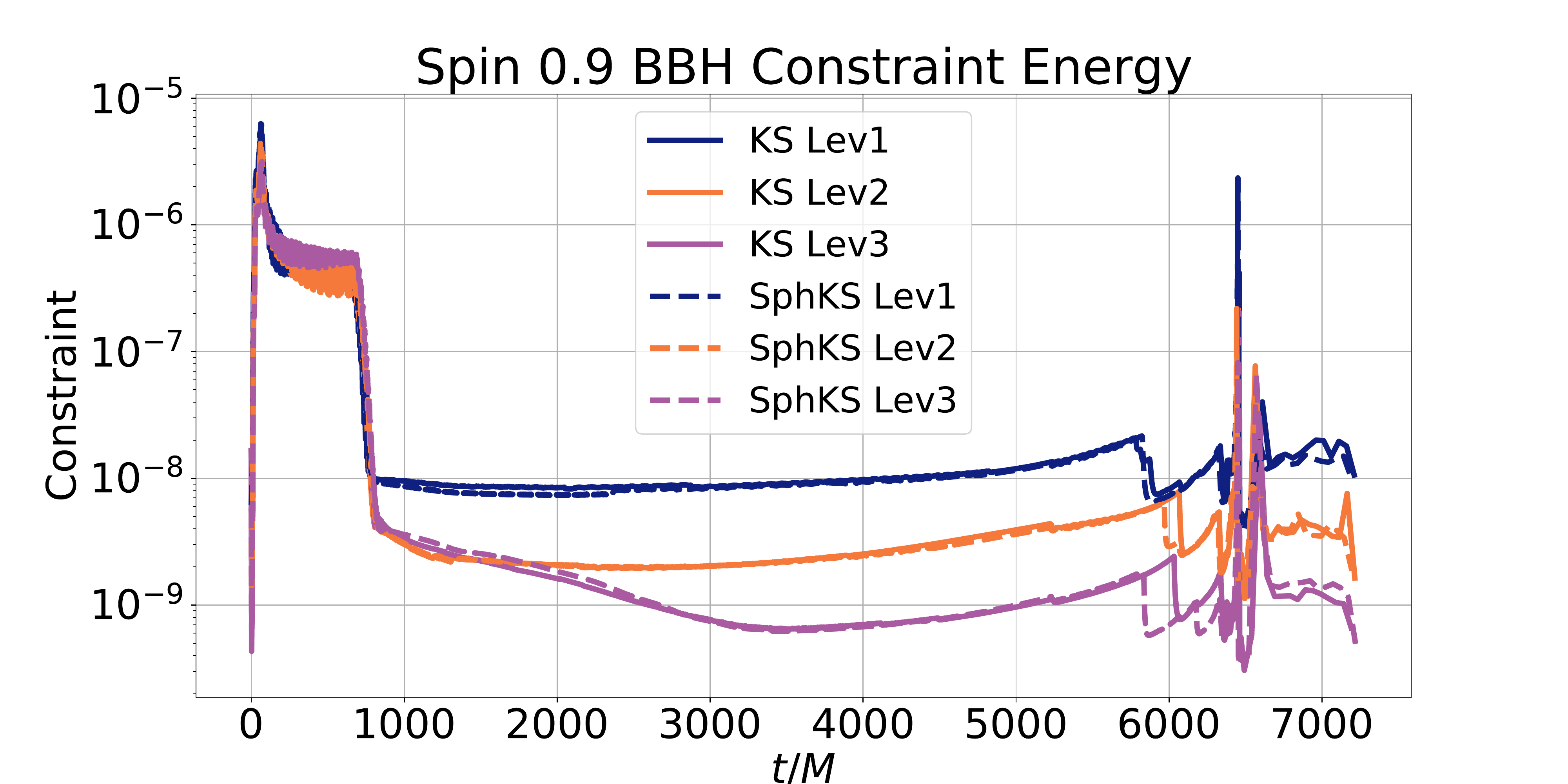}
  \includegraphics[width=0.49\linewidth]{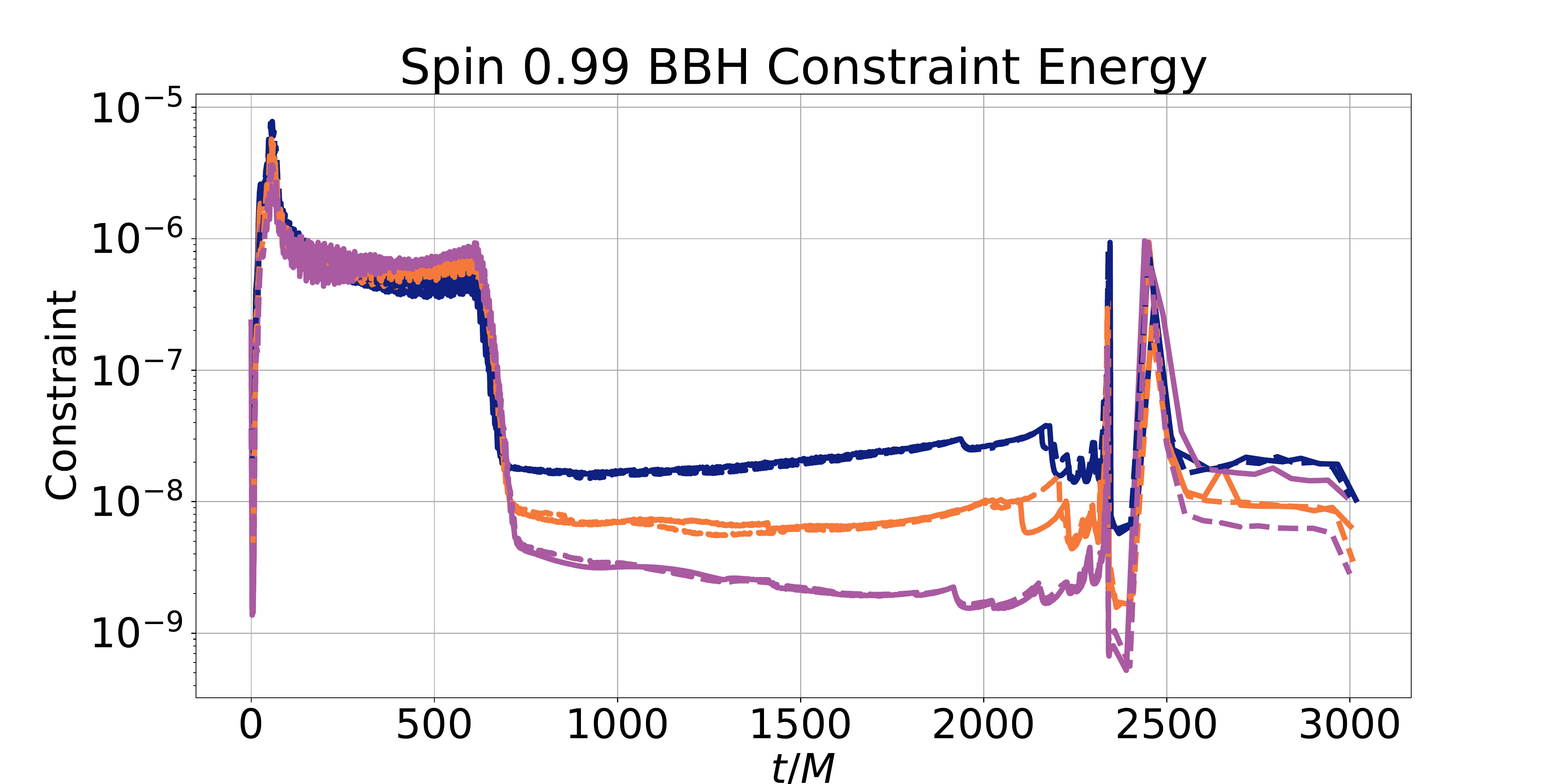}
  \caption{The left and right columns correspond to spin 0.9 and 0.99. The top
	row shows the ratio of CPU times ($T_{\mathrm{SphKS}}/T_{\mathrm{KS}}$)
	between the same Lev using the SphKS and KS initial data. In all cases, we see
	that initially the SphKS initial data is nearly 2 times faster than the KS
	initial data, but late in the simulation this ratio	gets closer to 1 since the
	majority of the simulation is	performed using the damped harmonic gauge. The
	bottom row shows the constraint	energy for the different simulations at the
	different resolutions. We see that in all cases the constraint violations
	between the SphKS and KS initial data	are nearly indistinguishable,
	demonstrating that the SphKS initial data can	achieve similar constraint
	violations as the KS initial data at significantly reduced computational
	cost.} 
  \label{fig:bbh_ghce_cpu}
\end{figure*}

\begin{figure*}[t]
  \centering \includegraphics[width=0.55\linewidth]{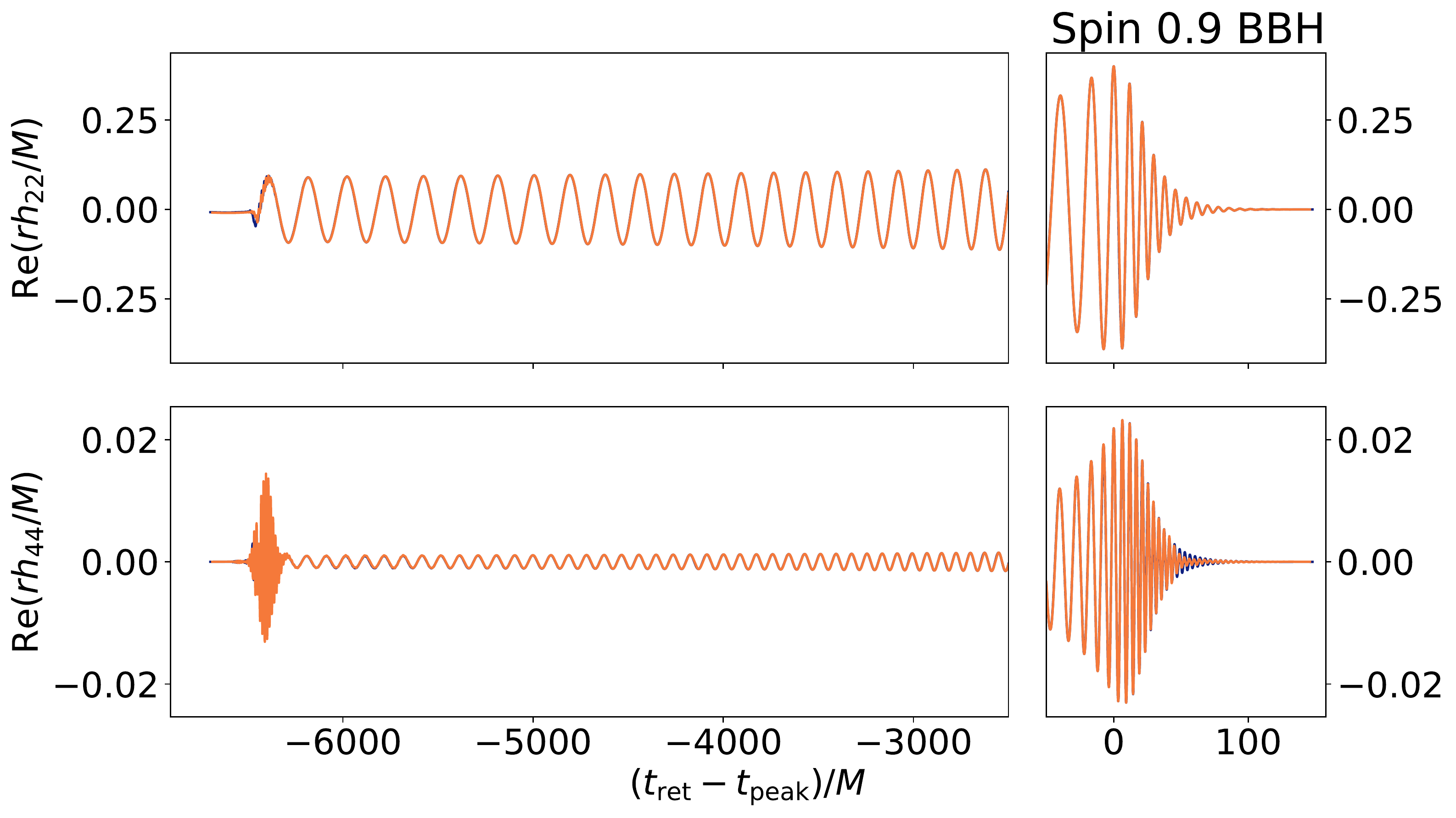}
  \includegraphics[width=0.29\linewidth]{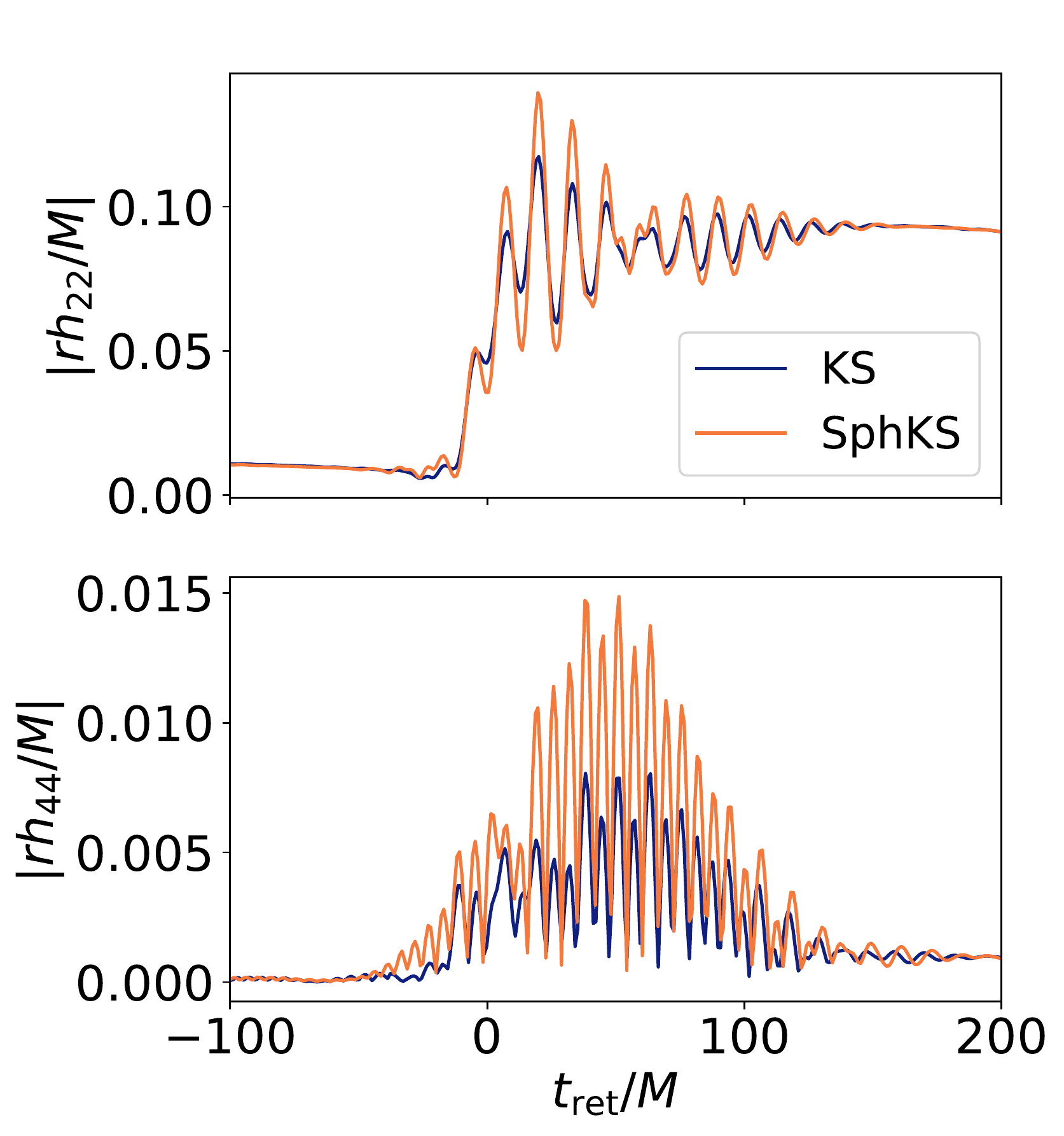}
  \includegraphics[width=0.55\linewidth]{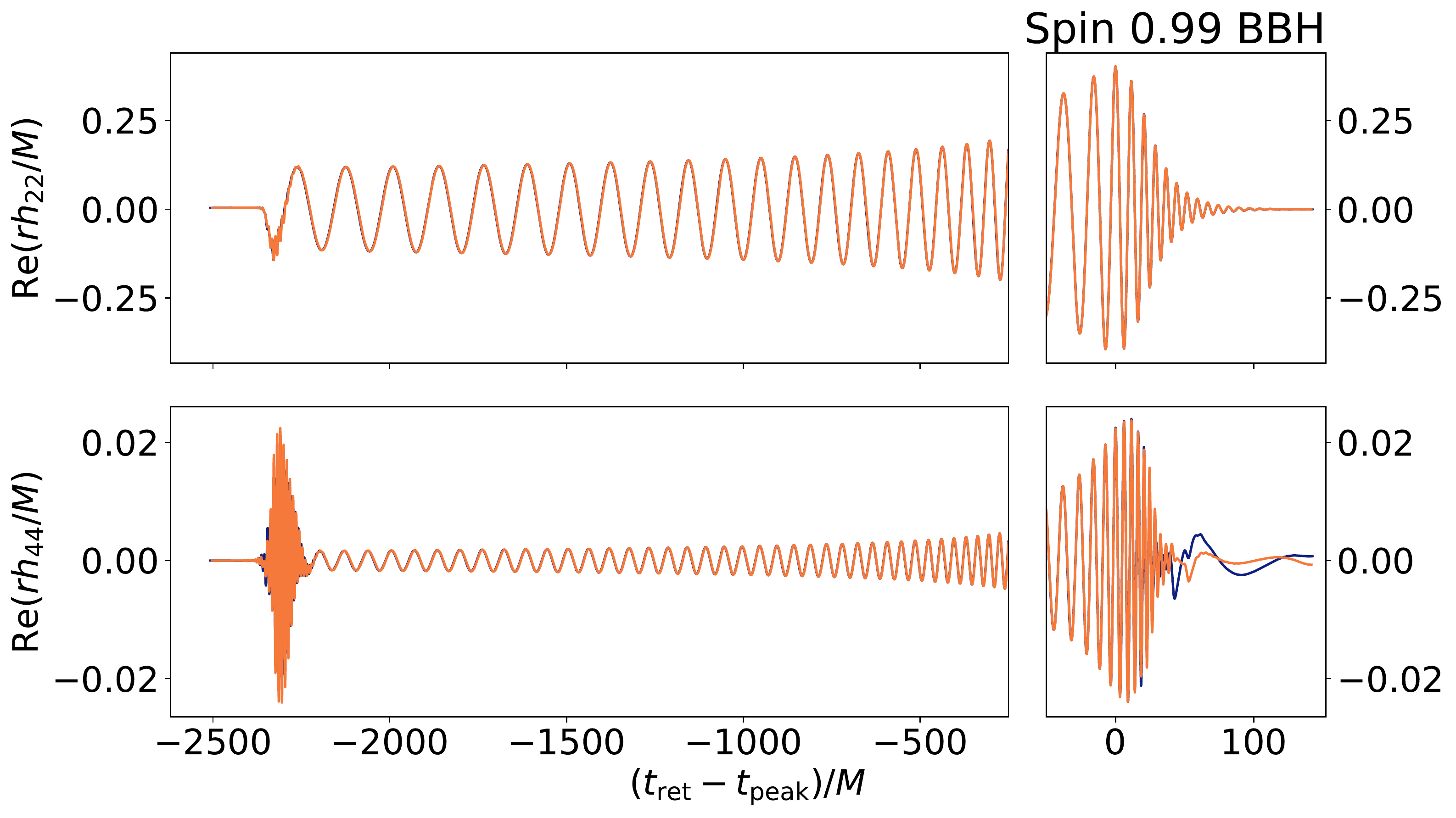}
  \includegraphics[width=0.29\linewidth]{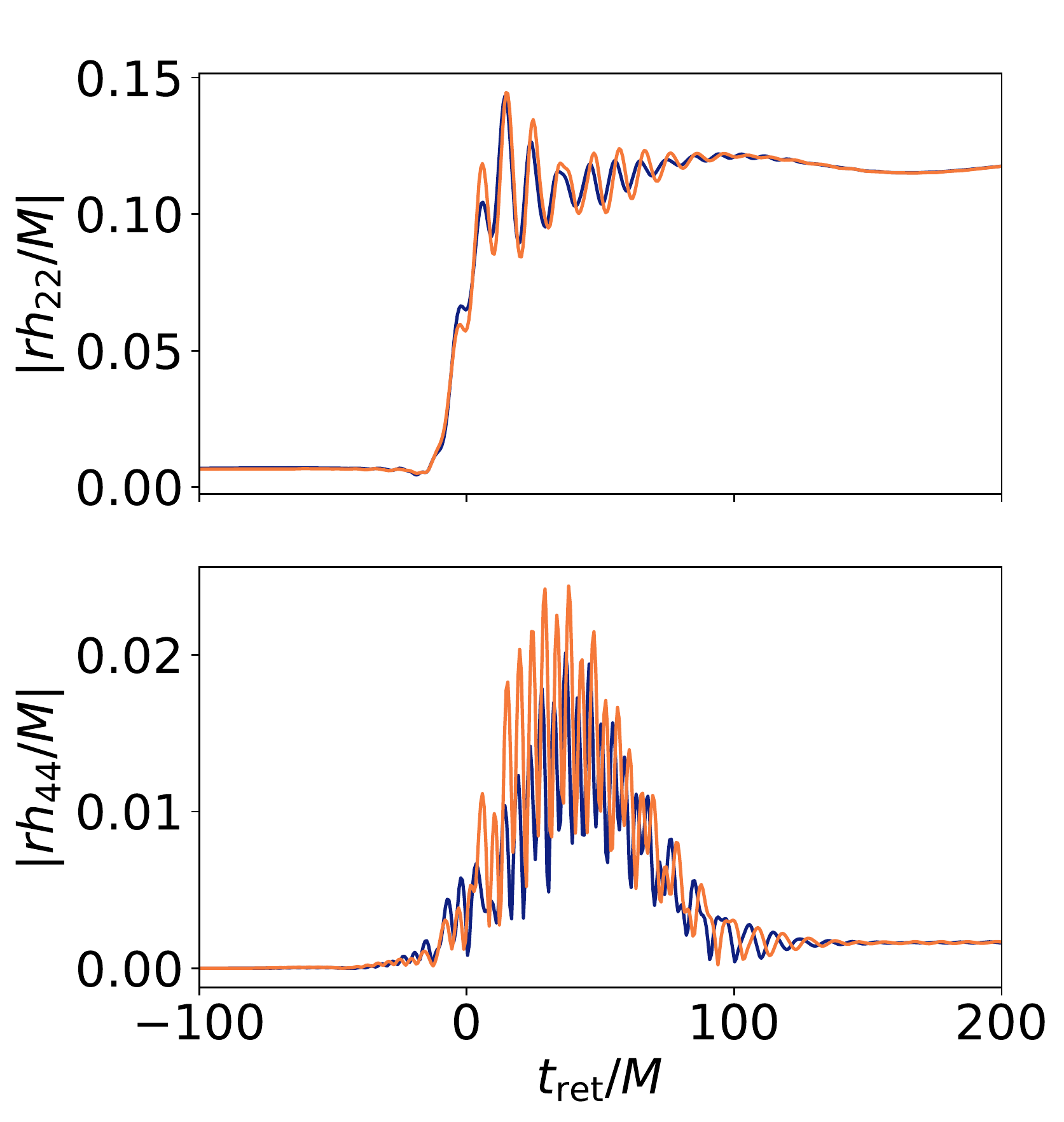}
  \caption{Strain $rh$ (extrapolated to
		$\mathscr{I}^+$) as a function of retarded time for the Lev-3 simulations.
		The top and bottom rows are for spin 0.9 and 0.99. The left column shows Re($rh_{22}$) and Re($rh_{44}$), while the right column shows $|rh_{22}|$ and $|rh_{44}|$. Waveforms in the left column are all time-shifted so that $|rh_{22}|$ (but not $|rh_{44}|$) reaches its maximum at time 0. They are	also phase-shifted so that both $rh_{22}$ and $rh_{44}$ are real at time 0.	Waveforms for KS and SphKS overlap well, except for the junk stage and tens of $M$ after merger in mode (4, 4). Waveforms in the right column are not time-shifted and only the junk parts are displayed. The amount of junk radiation for both gauges is comparable.} 
	\label{fig:bbh_rh}
\end{figure*}

We focus on the spin-0.9 and spin-0.99 simulations in this paper. Comparisons of CPU times and constraint energy [Eq.~\eqref{eqn:ghce}] between the two gauges for the spin-0.9 and spin-0.99 simulations are shown in Fig.~\ref{fig:bbh_ghce_cpu}. Comparison of waveforms for the spin-0.9 and	spin-0.99 Lev-3 simulations are shown in Fig.~\ref{fig:bbh_rh}. Additionally, the CPU times\footnote{Because the number of CPUs used may vary during a simulation, CPU time is a better measure of efficiency than wallclock time and we do not include wallclock time in this paper.} at the end of ringdown are recorded in Table~\ref{tbl:bbh_speed}. We will explain and analyze these figures and tables in greater detail. 

We do not show figures for the spin-0.95 case because the spin-0.95 and spin-0.99 simulations share the same qualitative behavior. However, we still provide the CPU times of the spin-0.95 case in Table~\ref{tbl:bbh_speed} to show the trend that using SphKS accelerates BBH simulations more significantly as the spin increases.

\begin{table}
	\caption{CPU times at the end of the six BBH simulations. $T_{\text{KS}}$ and
		$T_{\text{SphKS}}$ are the CPU times with the KS and SphKS initial data. We
		calculate the	ratio $T_{\text{SphKS}}/T_{\text{KS}}$ for each spin and Lev in the fifth column. A smaller ratio	means the SphKS simulation is more efficient than the KS simulation. We see improvements in all cases. The CPU time ratio ranges from $\sim$0.56 to $\sim$0.93. In the two most expensive runs, spin-0.95 Lev-3 and spin-0.99 Lev-3	runs, using the SphKS initial data is almost 2 times faster than using the KS initial data.} 
	\label{tbl:bbh_speed}
	\begin{ruledtabular}
		\begin{tabular}{ccccc}
			Spin & Lev & $T_{\text{KS}}$ [hr] & $T_{\text{SphKS}}$ [hr] & $T_{\text{SphKS}}/T_{\text{KS}}$ \\ 
			\hline \noalign{\vskip 1mm} 
			0.9 & 1 & 10582 & 8307 & 0.785 \\
			& 2 & 16957 & 14628 & 0.863 \\
			& 3 & 23369 & 21272 & 0.910 \\
			0.95 & 1 & 7318 & 6778 & 0.926 \\
			& 2 & 10273 & 9094 & 0.885 \\
			& 3 & 22697 & 13104 & 0.577 \\
			0.99 & 1 & 14651 & 13147 & 0.897 \\
			& 2 & 28764 & 18312 & 0.637 \\
			& 3 & 70709 & 39583 & 0.560 \\
		\end{tabular}
	\end{ruledtabular}
\end{table}

\subsubsection{Efficiency and constraint energy}

The top left panel of Fig.~\ref{fig:bbh_ghce_cpu} shows the CPU time ratio
$T_{\text{SphKS}}/T_{\text{KS}}$ as a function of simulation time for the
spin-0.9 BBH simulations, where $T_{\text{SphKS}}$ and $T_{\text{KS}}$ are the
CPU times of SphKS and KS runs. A ratio smaller than 1 means SphKS is more
efficient than KS. The smaller the ratio the more efficient the SphKS simulation
is. Since all curves in the top left panel are below 1, the performance of SphKS
runs is overall better. The ratio $T_{\text{SphKS}}/T_{\text{KS}}$ at the end of
the simulation (after ringdown) is listed in Table~\ref{tbl:bbh_speed}, together
with the CPU times. This CPU time ratio ranges from 0.79 to 0.91, which means a
significant improvement for a BBH simulation by switching to the SphKS initial
data.

The bottom left panel shows the constraint energy for both gauges. Solid lines
stand for KS while dashed lines for SphKS. Curves of the same resolution (Lev)
are plotted in the same color. We see that the constraint energy is similar for
the same Lev between the SphKS and KS initial data. This is because AMR adjusts
the target truncation error to control the constraint violations. What the top
and bottom panels together show is that the SphKS initial data allows us to
perform simulations with the same constraint violations at a reduced
computational cost. Furthermore, we observe exponential convergence of the
constraint energy between $t\sim700M$ and merger. Before $t\sim700M$, the curves
in different Levs overlap, and their values are much greater than the later
portion. This is because the AMR algorithm is disabled in the wave zone before
$t\sim700M$ to avoid using excessive computational resources to resolve the junk
radiation.

\subsubsection{Waveforms}
We extract the strain $h$ on multiple spherical surfaces of Euclidean radii $r$
and extrapolate $r h$ to $\mathscr{I}^+$ as a function of retarded time
$t_{\text{ret}}$ \cite{0905.3177, 1302.2919, 1409.4431, 1509.00862, 2010.15200}.
Note that $rh$ is always center-of-mass-corrected in this paper. We show the
$(l=2, m=2)$ and $(l=4, m=4)$ modes of $rh$, denoted as $rh_{22}$ and $rh_{44}$,
in the first row of Fig.~\ref{fig:bbh_rh}. Only the waveforms of the Lev-3
simulations are plotted. In each graph, the blue curve is data from the KS
simulation and the orange curve from the SphKS simulation.

The top left diagram of Fig.~\ref{fig:bbh_rh} shows the real part of $rh_{22}$
and $rh_{44}$. For clearer comparison, these waveforms are both time-shifted and
phase-shifted. Within the range of the retarded time $t_{\text{ret}}$, we choose
the point $t_{\text{peak}}$ at which $|rh_{22}|$ reaches its maximum and shift
the horizontal axis by $t_{\text{peak}}$. We then multiply each waveform by a
phase such that the waveform is real and positive at $t_{\text{peak}}$. In other
words, after time-shifting and phase-shifting, $|rh_{22}|$ is peaked (but not
necessarily $|rh_{44}|$) at $t=t_{\text{peak}}$, and both $rh_{22}$ and
$rh_{44}$ have zero phase at time $t_{\text{ret}}-t_{\text{peak}}=0$. This is
similar to the procedure in Ref.~\cite{1808.08228}. The waveforms of the KS and
SphKS simulations overlap very well after the junk radiation,
$t_{\text{ret}}\gtrsim700M$, except for some $\lesssim0.001$ deviations after
$t_{\text{ret}}-t_{\text{peak}}\sim40M$ during the ringdown phase.

We quantify the similarity between two waveforms by the \textit{mismatch}
$\mathcal{M}$,
\begin{align}
  \mathcal{M}(h_1,h_2) = 1 - \max_{\delta\phi,\delta t} \left[\frac{|\langle
  h_1|h_{2,\delta\phi,\delta t}\rangle|}{\sqrt{\langle h_1|h_1\rangle \langle
  h_{2,\delta\phi,\delta t}|h_{2,\delta\phi,\delta
  t}\rangle}}\right], \label{eqn:mismatch}
\end{align}
where $h_{2,\delta\phi,\delta t} = e^{i\delta\phi} h_2(t+\delta t)$ and
$h_1, h_2$ are waveforms in a specific mode. $\delta\phi$ and $\delta t$ are
parameters in phase- and time-shifting to maximize the overlap between two
waveforms. The inner product $\langle \cdot|\cdot\rangle$ is defined as
\begin{align}
  \langle f|g\rangle = \int_{t_i}^{t_f} f(t) g^*(t) dt,
\end{align}
where * denotes complex conjugation. The mismatch is calculated for each mode
($h_{22}$ or $h_{44}$) over the time domain, unlike Ref.~\cite{1904.04831},
which considers the strain $h$ before mode decomposition and calculates the
inner product over the frequency domain. We choose $t_i$ to be $700M$ after the
earliest time in KS Lev-3 waveform and $t_f = t_\mathrm{peak} + 50M$. For both
the $(2, 2)$ and $(4, 4)$ modes, the mismatch between KS Lev-2 and KS Lev-3 are
at the same level as the mismatch between KS Lev-3 and SphKS Lev-3.\footnote{For
	the $(2, 2)$ mode, the mismatch between KS Lev-2 and KS Lev-3 is $6.44\times
	10^{-6}$, while the mismatch between KS Lev-3 and SphKS Lev-3 is $9.84\times
	10^{-7}$. For the $(4, 4)$ mode, the mismatch between KS Lev-2 and KS Lev-3 is
	$6.72\times 10^{-5}$, while the one between KS Lev-3 and SphKS Lev-3 is
	$1.61\times 10^{-4}$. } Thus, the waveforms after junk radiation passes are in
good agreement between KS and SphKS.

The high-frequency fluctuation within $t_{\text{ret}}\lesssim700M$ is the
transient gravitational perturbation called \textit{junk radiation}. The origin
of the junk radiation is the initial data not representing the true spacetime
snapshot of a BBH system in quasi-equilibrium. The top right diagram of
Fig.~\ref{fig:bbh_rh} shows $|rh_{22}|$ and $|rh_{44}|$ during the junk phase.
This waveform is not time-shifted because we only care about the qualitative
comparison of junk radiation between the two gauges. Note that the retarded time
$t_{\text{ret}}$ can extend to negative values, and the waveform on this
negative time axis corresponds to perturbations in the wave zone initial data.
Both the KS and SphKS gauges produce roughly the same amount of junk radiation
in the $(2, 2)$ mode, while SphKS produces more than KS in the $(4, 4)$ mode.

\subsection{Spin-0.99 binary-black-hole simulations using spherical Kerr-Schild
  initial data} \label{sec:spin0.99bbh}

We omit a discussion of the spin-0.95 BBH data and jump directly to the spin-0.99 BBH case because the efficiency and waveform comparisons are very similar.

\subsubsection{Efficiency, constraint energy and waveforms}
The top right panel of Fig.~\ref{fig:bbh_ghce_cpu} shows the CPU time ratio of
the SphKS initial data to the KS initial data, while the bottom right panel of
Fig.~\ref{fig:bbh_ghce_cpu} shows the constraint energy at different resolutions
for the two gauges. The curves and axes are labeled the same as for the spin-0.9
BBH case. Overall, the behavior of the CPU time ratio and constraint violations
are similar to what we observed for the spin-0.9 simulations. Specifically, the
constraint violations for the SphKS and KS initial data are very similar, while
the simulations using SphKS initial data are cheaper than those using KS initial
data. The Lev-3 spin-0.99 SphKS simulation is almost two times faster than the
KS simulation.

The bottom row in Fig.~\ref{fig:bbh_rh} shows the waveforms of the spin-0.99 BBH
simulations for the KS (blue) and SphKS (orange) initial data. The waveforms of
the two gauges overlap well in both modes $(2, 2)$ and $(4, 4)$, except for some
deviation at later times
during ringdown. The mismatch [Eq.~\eqref{eqn:mismatch}]
between KS Lev-3 and SphKS Lev-3 is also at the same order as between KS Lev-2
and Lev-3 for each mode. We note that both gauges have roughly the same amount (but not the exact same form)
of junk radiation in both the $(2, 2)$ and $(4, 4)$ modes, in contrast with the
spin-0.9 case, where the SphKS gauge $(4,4)$ mode had more junk radiation. The
waveforms for both the SphKS and KS initial data being very similar and the
SphKS initial data simulation being nearly twice as fast for higher resolutions
demonstrate the advantage of using SphKS initial data for accurate and efficient
high-spin BBH simulations.

\subsubsection{Apparent horizon analysis}

SpEC decomposes the computational domain into multiple subdomains
(Sec.~\ref{sec:domain}). There is an innermost spherical shell subdomain that
encircles each BH and contains the BH's apparent horizon (AH). The shape of the
AH needs to be resolved, so we expect more spherical AHs to require lower
resolutions. For concreteness, we will focus on the AHs of the Lev-3 spin-0.99
BBH simulations.

\begin{figure}[t]
  \centering \includegraphics[width=0.49\linewidth]{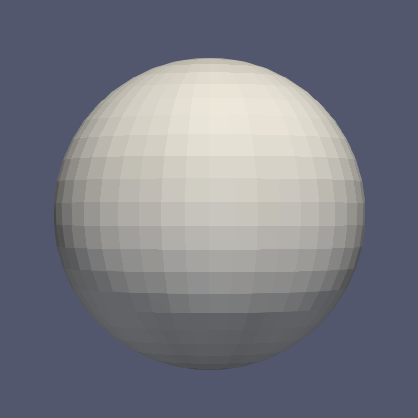}
  \includegraphics[width=0.49\linewidth]{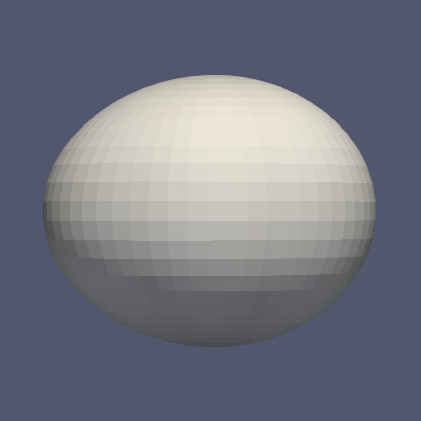}
  \caption{Apparent horizons of a progenitor-BH in the SphKS spin-0.99 Lev-3 BBH
    system, at $t=0M$ (left) and $t=50M$ (right). The apparent horizon is
    spherical at $t=0M$, which is a key feature of the SphKS gauge. The
    transition to damped harmonic gauge is mostly complete by $t=50M$, when we
    can see that the horizon is no longer spherical.}
  \label{fig:spin0.99_bbh_AH}
\end{figure}

\begin{figure}[t]
  \centering \includegraphics[width=\linewidth]{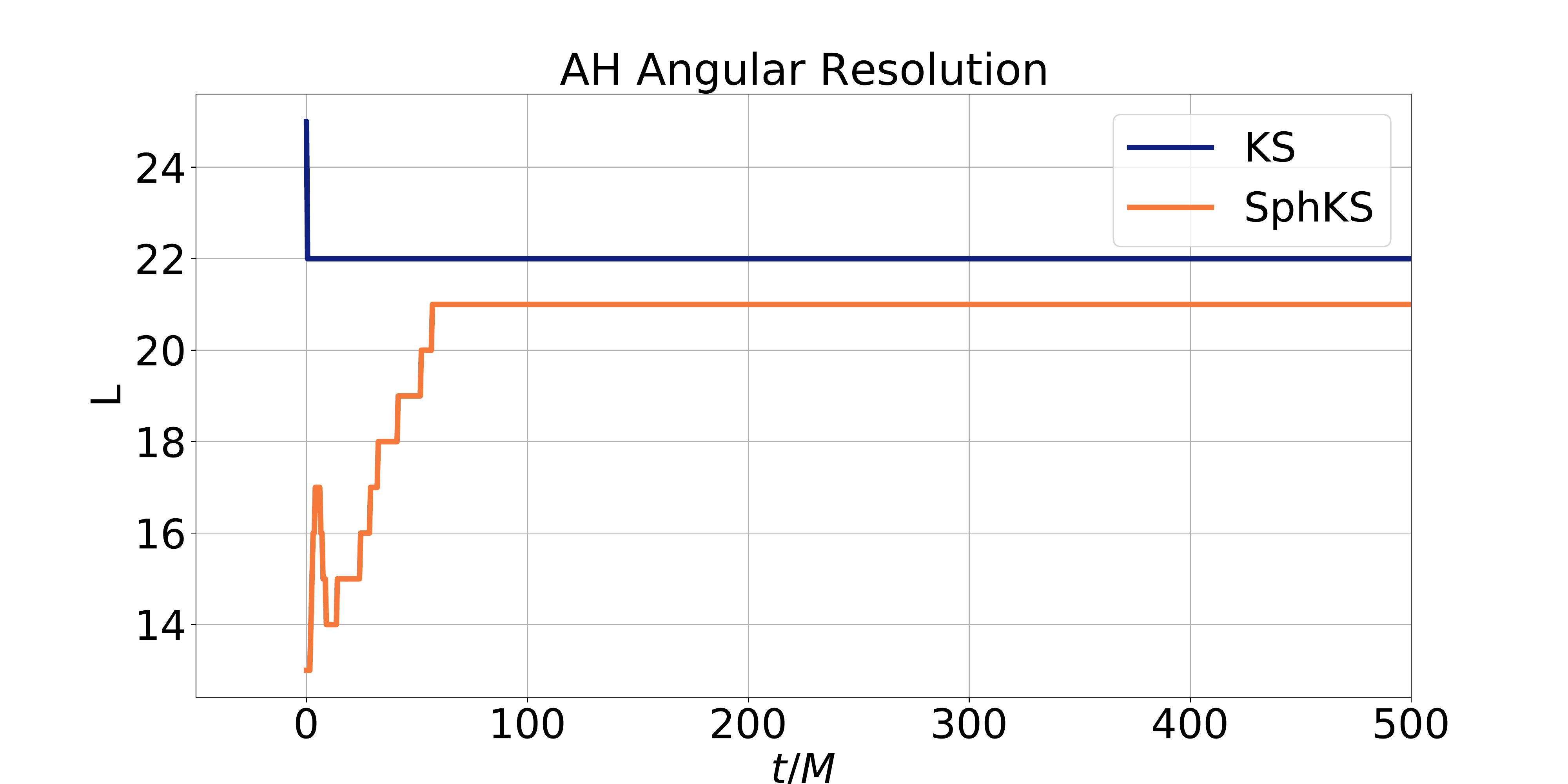}
  \includegraphics[width=\linewidth]{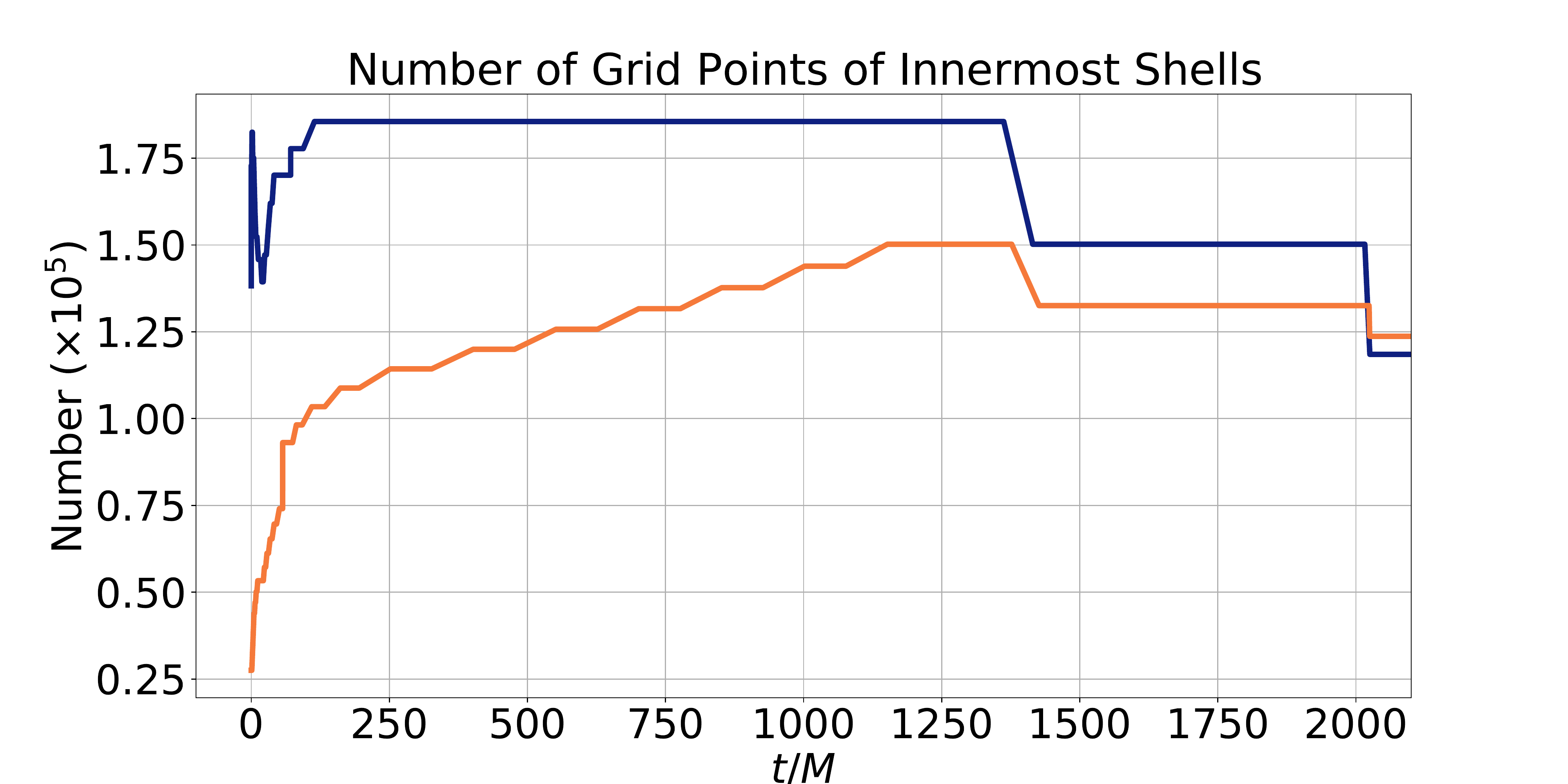}
  \caption{The top graph depicts the angular resolution $L$ used by the AH
    finder. $L$ in the KS run stays constant immediately after $t=0M$,
    suggesting that the AHs in the DH and KS gauges are similar. In the SphKS
    case, $L$ starts at a relatively low value and then increases to a constant
    during the DH gauge transition. The bottom graph records the total number of
    grid points in the innermost spheres surrounding the BHs.}
  \label{fig:spin0.99_bbh_sphere0}
\end{figure}

Figure~\ref{fig:spin0.99_bbh_AH} shows the AH profile of a progenitor-BH at two
different times in the SphKS run. The left picture is at the beginning of
evolution ($t=0M$) and clearly shows that the horizon is spherical in the SphKS
gauge. At $t=0M$, the simulation starts to undergo a smooth transition from the
quasi-equilibrium gauge to damped harmonic (DH) gauge, with the temporal width
$w=50M$ (Sec.~\ref{sec:evolution}). The right picture shows the AH at $t=50M$,
at which point
the AH is already non-spherical. In addition to the images of the AH,
we record the angular resolution $L$ used to construct AHs by the AH finder
\cite{1211.6079}, in the top panel of Fig.~\ref{fig:spin0.99_bbh_sphere0}. The
graph shows the angular resolution $L$ of Lev-3 runs in both gauges for
$t<500M$. Because the AH in KS gauge starts as a distorted spheroid, the angular
resolution $L$ in the KS simulation stays at $L=22$ immediately after the start
and throughout the DH gauge transition. This suggests that the AH in the DH
gauge is close to the spheroidal AH in the KS gauge. In the SphKS gauge, $L$
starts from a relatively low value (13) because of the spherical shape of the AH
in the initial data and then climbs to a constant value throughout the DH gauge
transition.  This behavior of $L$ matches our expectation.

\begin{figure*}[t]
	\centering \includegraphics[width=0.49\linewidth]{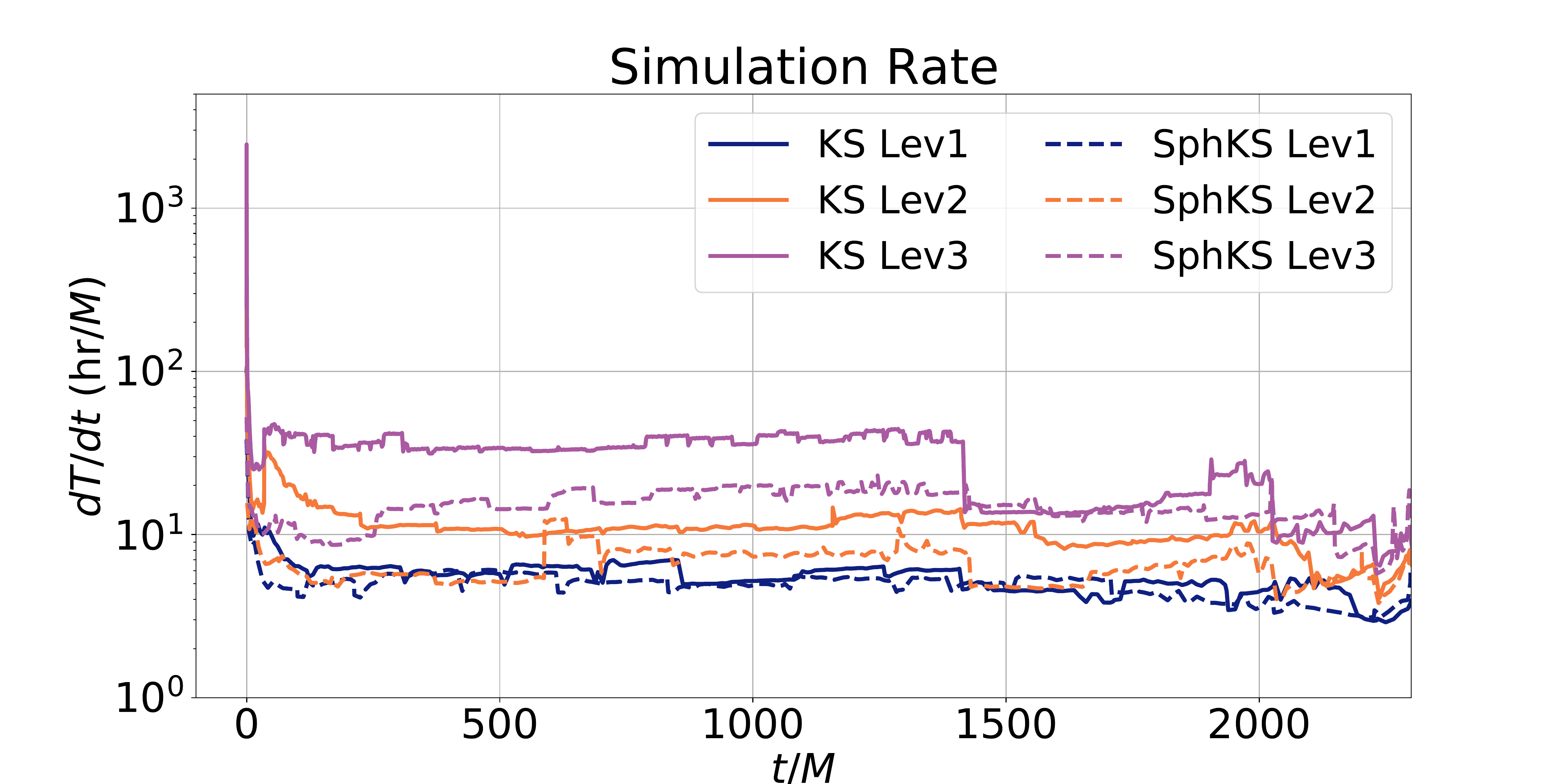}
	\includegraphics[width=0.49\linewidth]{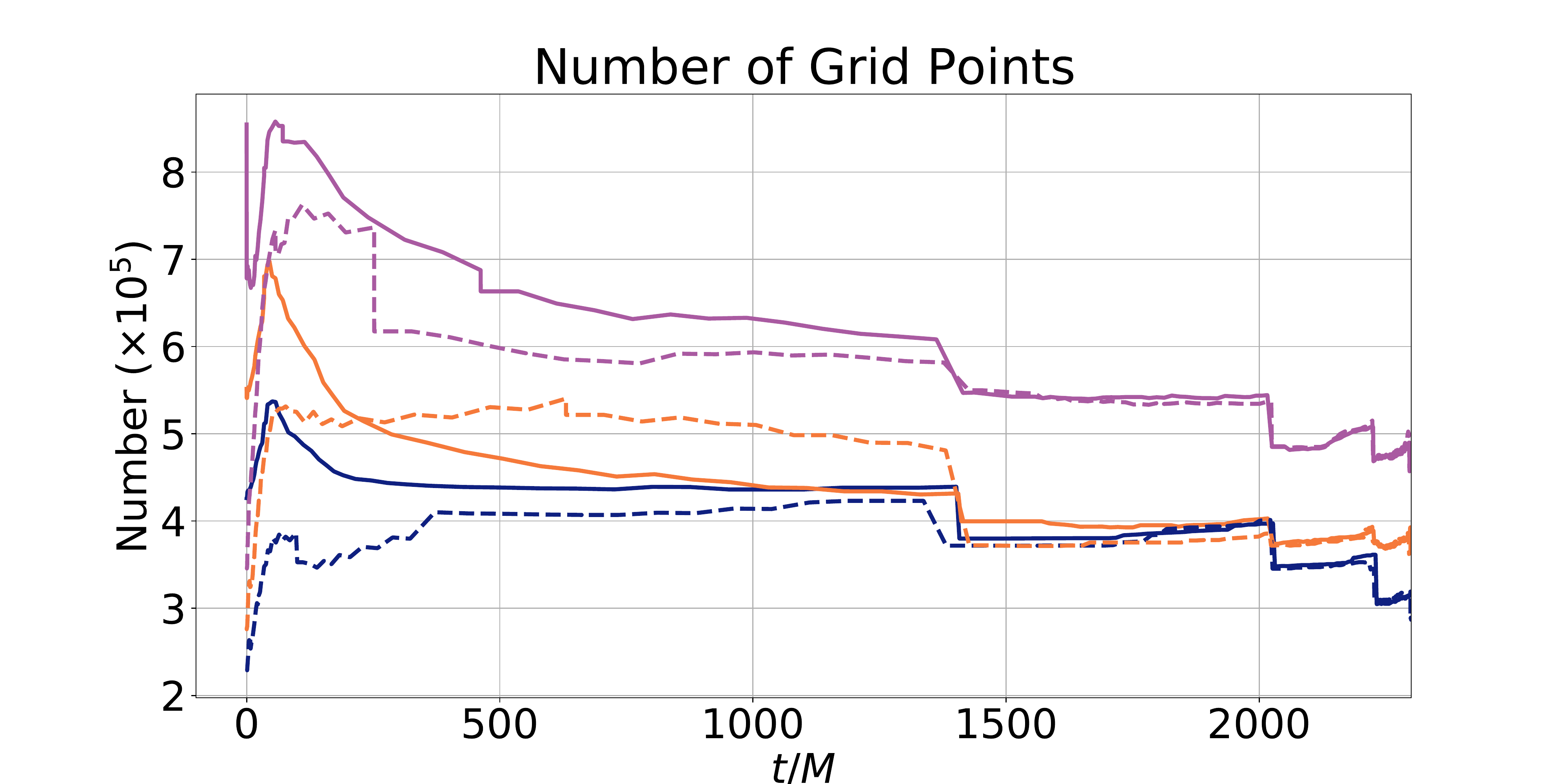}
	\includegraphics[width=0.49\linewidth]{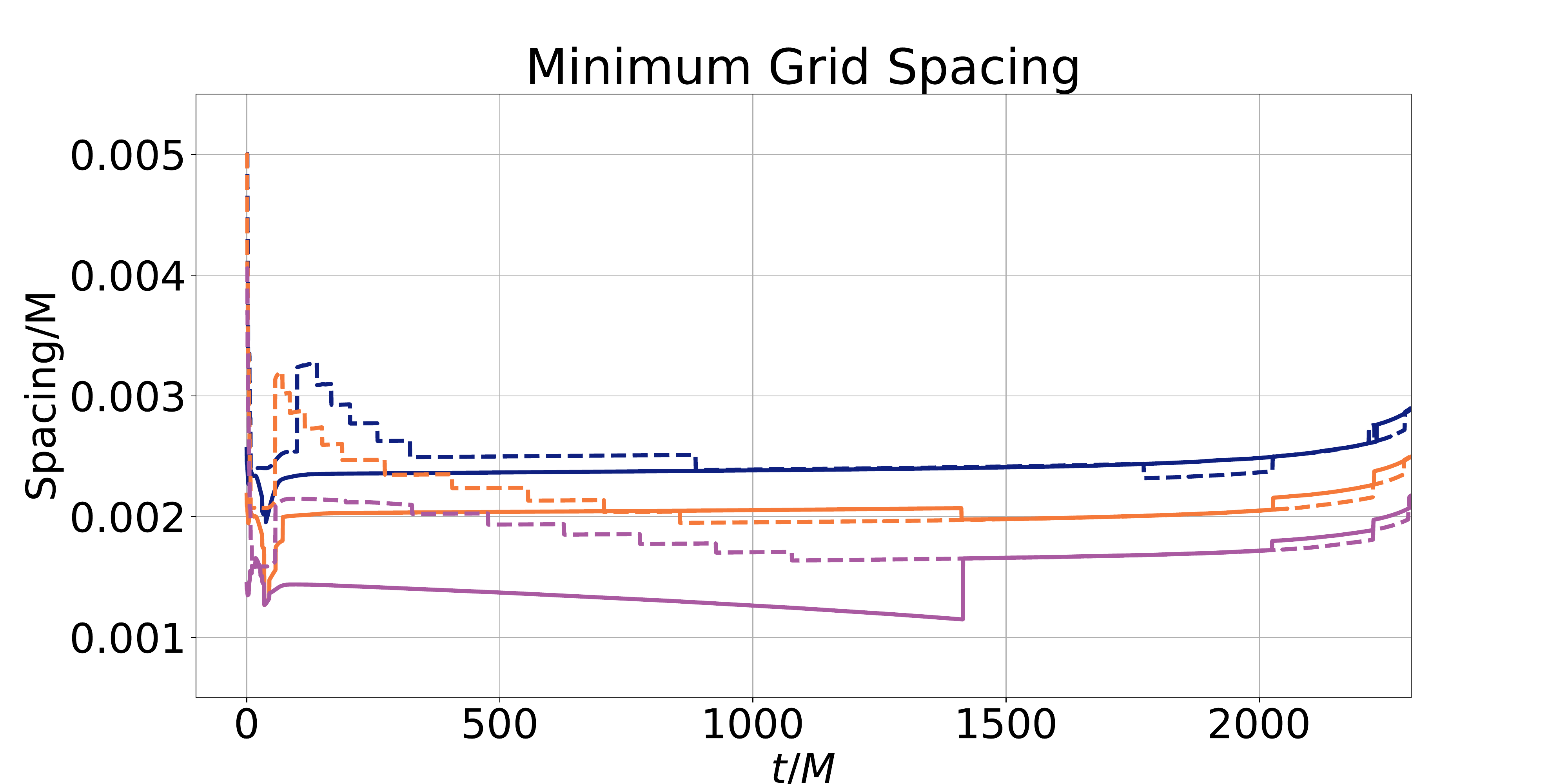}
	\includegraphics[width=0.49\linewidth]{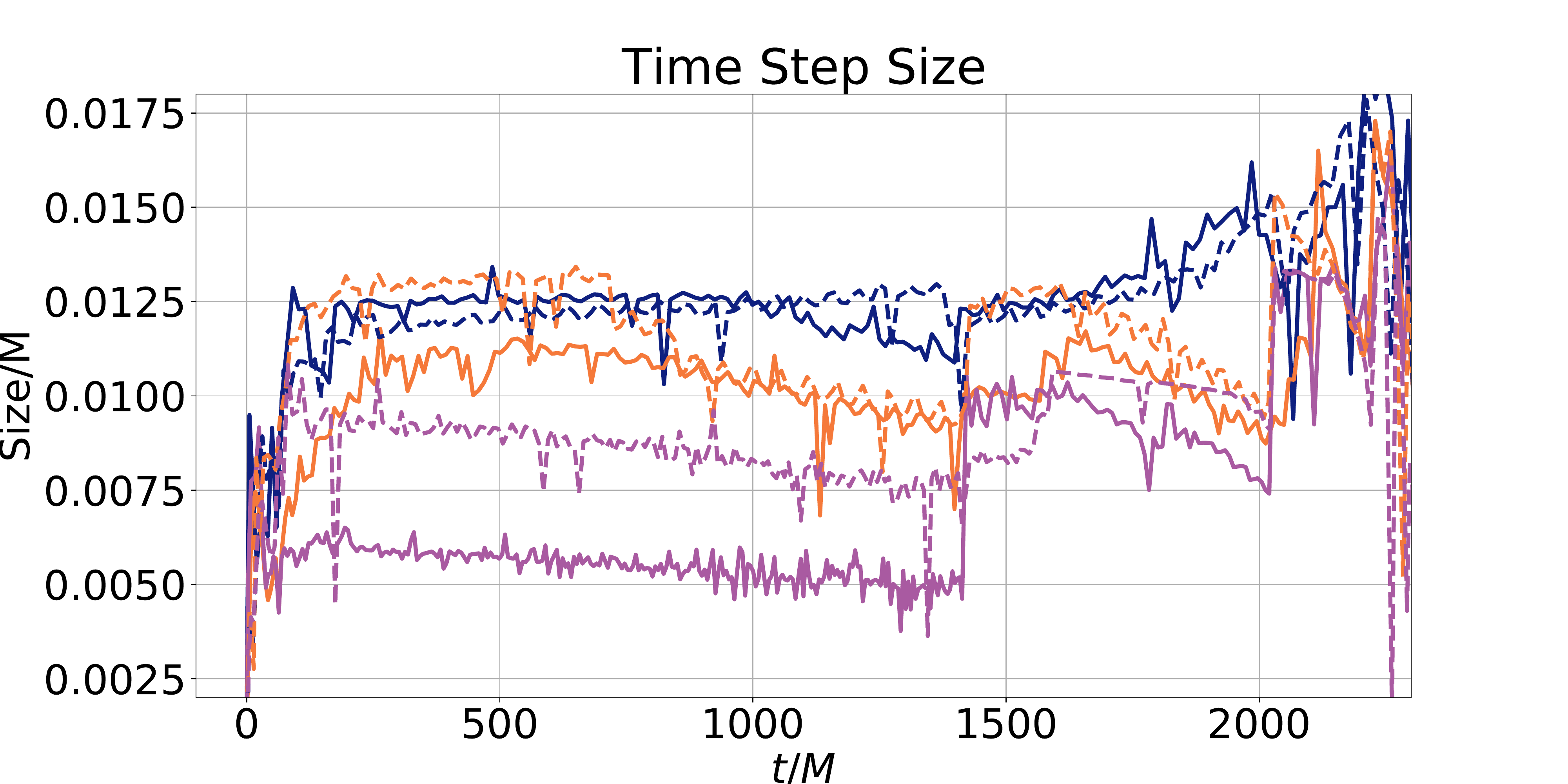}
	\caption{Four quantities related to the speed-up in the spin-0.99 BBH
		simulations. The top left graph shows the simulation rate $dT/dt$, and we
		see that the speed-up occurs throughout the evolution. The top right graph
		implies that the difference in the total number of grid points between two
		gauges cannot fully explain the speed-up. The bottom two graphs indicate that
		the narrower grid spacing in the KS simulations is a key factor making them
		slower than the SphKS simulations. The decreased grid spacing requires a
		smaller time step size and so the simulation progresses more slowly. Note
		that the abrupt change at $\sim1400M$ in Lev-3 is shared among all four
		graphs and the bottom panel of Fig.~\ref{fig:spin0.99_bbh_sphere0} and is
		caused by an AMR decision.}
	\label{fig:spin0.99_bbh_speed_analysis}
\end{figure*}

Although the AHs quickly lose their spherical shape, we have found that the
SphKS increases the speed of the simulation far beyond $t\sim50M$ (see the top
right panel of Fig.~\ref{fig:bbh_ghce_cpu} and also the top left panel of
Fig.~\ref{fig:spin0.99_bbh_speed_analysis}). In the bottom panel of
Fig.~\ref{fig:spin0.99_bbh_sphere0} we show the number of grid points in the
innermost shell surrounding each BH. We see that even though the angular
resolution used by the AH finder is nearly identical once the transition to DH
gauge is complete, the number of points used near the BHs is significantly lower
for the SphKS initial data all the way to $t\sim2000M$.

\subsubsection{Speed-up analysis} \label{sec:spin0.99bbh_analysis_speed-up}

Speeding-up the simulations significantly while keeping constraint violations
and waveforms almost the same is encouraging. In this section we identify the
contributions that lead to such a large speed-up. It is almost impossible to
quantitatively decompose the speed-up into various algorithms' contributions, but
we may still obtain some qualitative insight from several relevant
diagnostics of the evolution.

Figure~\ref{fig:spin0.99_bbh_speed_analysis} shows four such quantities. They are
the simulation rate ($dT/dt$, i.e., the derivative of CPU time with respect to
simulation time), total number of grid points, minimum grid spacing, and time
step size.  There are six curves in each graph, representing various Levs and
gauges. Since most of the speed-up occurs before the merger ($\sim2340M$), we
restrict our plots to $0\leq t\leq 2300M$.

The earlier graph of CPU time ratio versus simulation time in
Fig.~\ref{fig:bbh_ghce_cpu} provides an overall comparison of computational
cost. By contrast, an instantaneous comparison of efficiency between gauges can
be quantified by the derivative $dT/dt$, as shown in the top left panel of
Fig.~\ref{fig:spin0.99_bbh_speed_analysis}. A slower simulation results in a
higher curve in this graph, since more CPU time is required to compute a unit of
simulation time. Given a fixed Lev, the curve for KS almost always lies above
the one for SphKS, so a SphKS run is faster than a KS run not only collectively,
but also at almost every moment. The difference in $dT/dt$ between gauges is
larger as Lev increases, meaning that the SphKS gauge is especially useful if
both high spin and high accuracy are desired, as seen previously in
Fig.~\ref{fig:bbh_ghce_cpu}. We note that along the solid purple curve (KS
Lev-3) in Fig.~\ref{fig:spin0.99_bbh_speed_analysis} the value of $dT/dt$ stays
nearly constant after the beginning, but then plummets to the level of the
dashed purple curve (SphKS Lev-3) at $\sim1400M$. This drop is related to the
AMR algorithm (Sec.~\ref{sec:domain}) rearranging the shells near the BHs.
Attempts to replicate this domain configuration at earlier times led to unstable
simulations. We will continue observing this behavior at $1400M$ throughout the
next several graphs.

To better understand the source of the speed-up that the SphKS initial data
provides, let us first look at the number of grid points used by the SphKS and
KS simulations as a function of time. The top right panel of
Fig.~\ref{fig:spin0.99_bbh_speed_analysis} shows the number of grid points as a
function of time for both gauges and all three resolutions. While the Lev-1 and
Lev-3 SphKS simulations have fewer grid points than KS, the trends of their
curves do not match the $dT/dt$ curves. For example, the number of grid points
for the Lev-3 simulations approach each other before $t\sim250M$, while $dT/dt$
is still much smaller for the SphKS simulation. More surprisingly, the Lev-2
SphKS simulation uses more total grid points than the Lev-2 KS simulation but
still has a smaller $dT/dt$. These differing trends between simulation rate and
number of grid points suggests that there is another major contributing factor
responsible for the observed speed-up.

Next we look at the Courant–Friedrichs–Lewy condition
\cite{Courant1_skip}, by which the time step size is adjusted according to the
spacing between grid points. If the spacing is narrower, then the time step size
must be smaller, making the simulation slower. We plot the minimum grid spacing
in the bottom left and the time step size in the bottom right panel of
Fig.~\ref{fig:spin0.99_bbh_speed_analysis}. We see that the minimum grid spacing
for SphKS is larger than for KS in the first several hundred $M$, except for the
region $t\lesssim40M$. However, they both reach the same level later in the
evolution for all Levs. The bottom right panel of
Fig.~\ref{fig:spin0.99_bbh_speed_analysis} shows that the SphKS time step size
is generally larger than the KS time step. Again, the difference is also more
noticeable in the first several hundred $M$.

Focusing on the Lev-3 curves we see that SphKS has a larger minimum grid
spacing, a larger time step size and faster simulation rate than KS in
$40M<t<1400M$. We also check that the minimum grid spacing is always located in
the innermost shells near BHs in this time range for both KS and SphKS
simulations. Around $1400M$ in the KS simulation AMR changes the grid, resulting
in fewer grid points in the innermost shells (see
Fig.~\ref{fig:spin0.99_bbh_sphere0}) and a minimum grid spacing, and a time step
comparable to the SphKS case. At this point, the SphKS and KS simulations also
have approximately the same $dT/dt$. This leads us to conclude that the more
sparse distribution of grid points near the BHs in a simulation with SphKS
initial data is the major contributing factor to the speed-up.

\subsection{Binary-black-hole simulations using wide Kerr-Schild initial
  data} \label{sec:spin0.9bbh_wks}

We evolve a 25-orbit, non-precessing, non-eccentric, equal-mass, spin-0.9 BBH
system using the wide Kerr-Schild (WKS) gauge (Sec.~\ref{sec:wks}) with $b=0.95$
and $b=0.9$ (recall that $b=1$ corresponds to the SphKS gauge). The parameters
of initial setup are listed in Table~\ref{tbl:spin0.9_bbh_wks_param}. Their
values are close to the previous spin-0.9 SphKS BBH run, so we compare these three simulations together.

\begin{table}
	\caption{Parameters of three BBH simulations using WKS initial data with
		$b\in\{1,0.95,0.9\}$. Note that WKS of $b=1$ is equivalent to SphKS. In all three simulations, the mass ratio is 1, $\chi_{A,B} = (0, 0, 0.9)$, and the number of orbits is about 25. See Table~\ref{tbl:bbh_param} for definitions of parameters.}
	\label{tbl:spin0.9_bbh_wks_param}
	\begin{ruledtabular}
		\begin{tabular}{ccccc}
			Initial gauge & $D_0$ $[M]$ & $\Omega_0$ & $\dot{a}_0$ & $e$ \\ 
			\hline \noalign{\vskip 1mm} 
			SphKS & 15.450 & 0.0142 & 4.53$\times 10^{-4}$ & $\sim0.0005$ \\
			WKS b0.95 & 15.452 & 0.0142 & 4.56$\times 10^{-4}$ & $\sim0.0003$ \\
			WKS b0.9 & 15.455 & 0.0142 & 4.51$\times 10^{-4}$ & $\sim0.0003$ \\
		\end{tabular}
	\end{ruledtabular}
\end{table}

We find that the WKS simulations share most of the properties of the SphKS
simulations. Namely, there is no consistent improvement in either CPU efficiency
or constraint energy by switching from SphKS to WKS. Both
the strain modes Re$(rh_{22})$ and Re$(rh_{44})$ from different gauges overlap
well. However, the amount of junk radiation significantly increases when
$b\ne1$. We find that the junk is approximately doubled when $b=0.9$ compared to
$b=1$. Therefore, we do not recommend the use of WKS for
evolutions of high-spin BBHs.

\subsection{Binary-black-hole simulations with delayed evolution gauge
  transition} \label{sec:spin0.9bbh_delay}

In this section, we simulate BBH systems in SphKS where we delay the transition
from the initial spherical gauge to damped harmonic (DH) gauge with the hope
that this further improves efficiency. We choose $t_0=4000M$ and
$w\in\{50M,100M,400M\}$ [see Eq.~\eqref{eqn:time_transition} for the definitions
of $t_0$ and $w$]. All of these cases share the same initial parameters with the
non-delayed-transition SphKS spin-0.9 run in Table~\ref{tbl:bbh_param}. All runs
have negligible eccentricity ($e<0.0007$), and we only focus on the highest
resolution, Lev-3.

\subsubsection{Efficiency and constraint energy}

\begin{figure}[t]
  \centering
  \includegraphics[width=\linewidth]{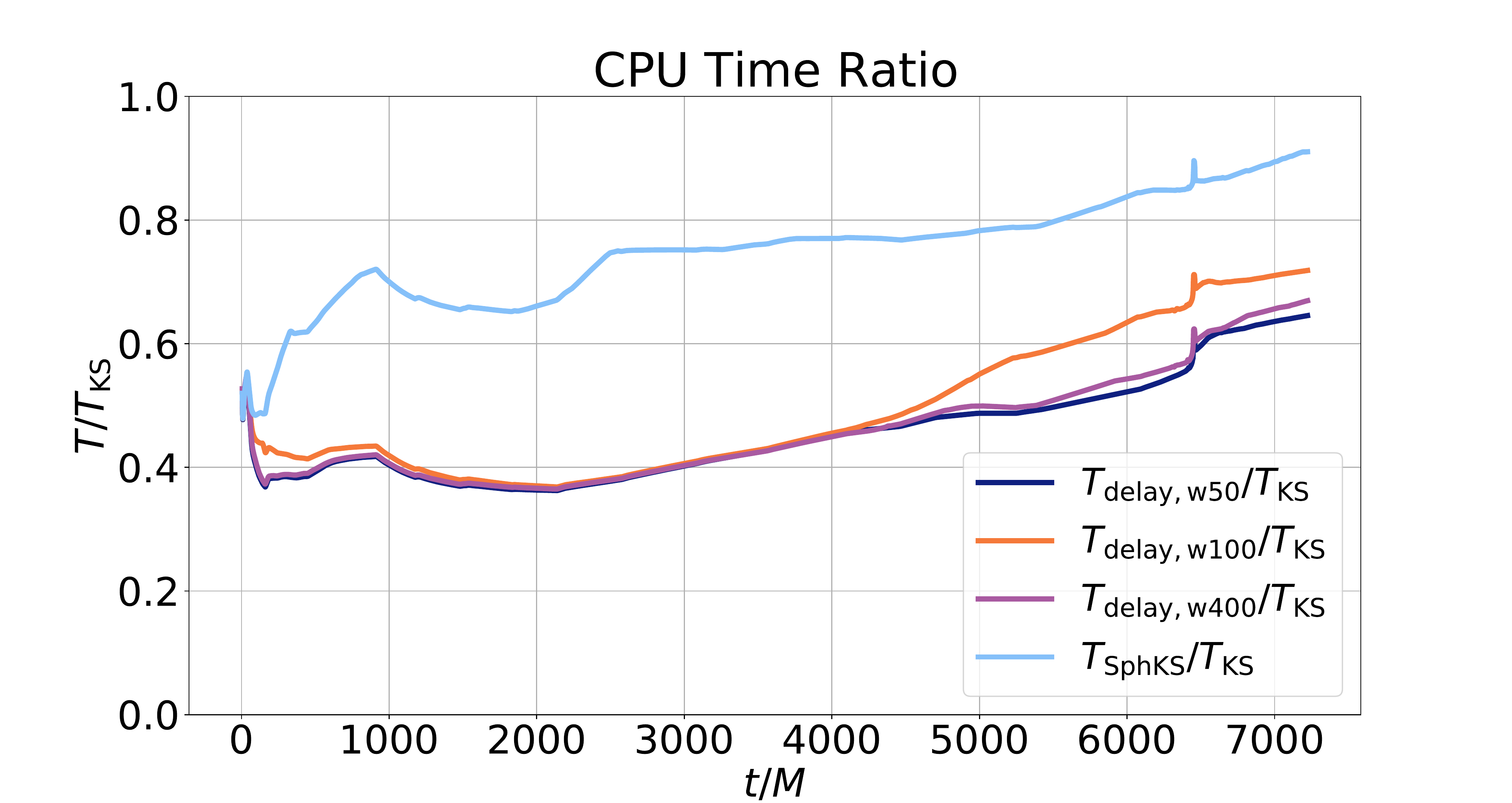}
  \includegraphics[width=\linewidth]{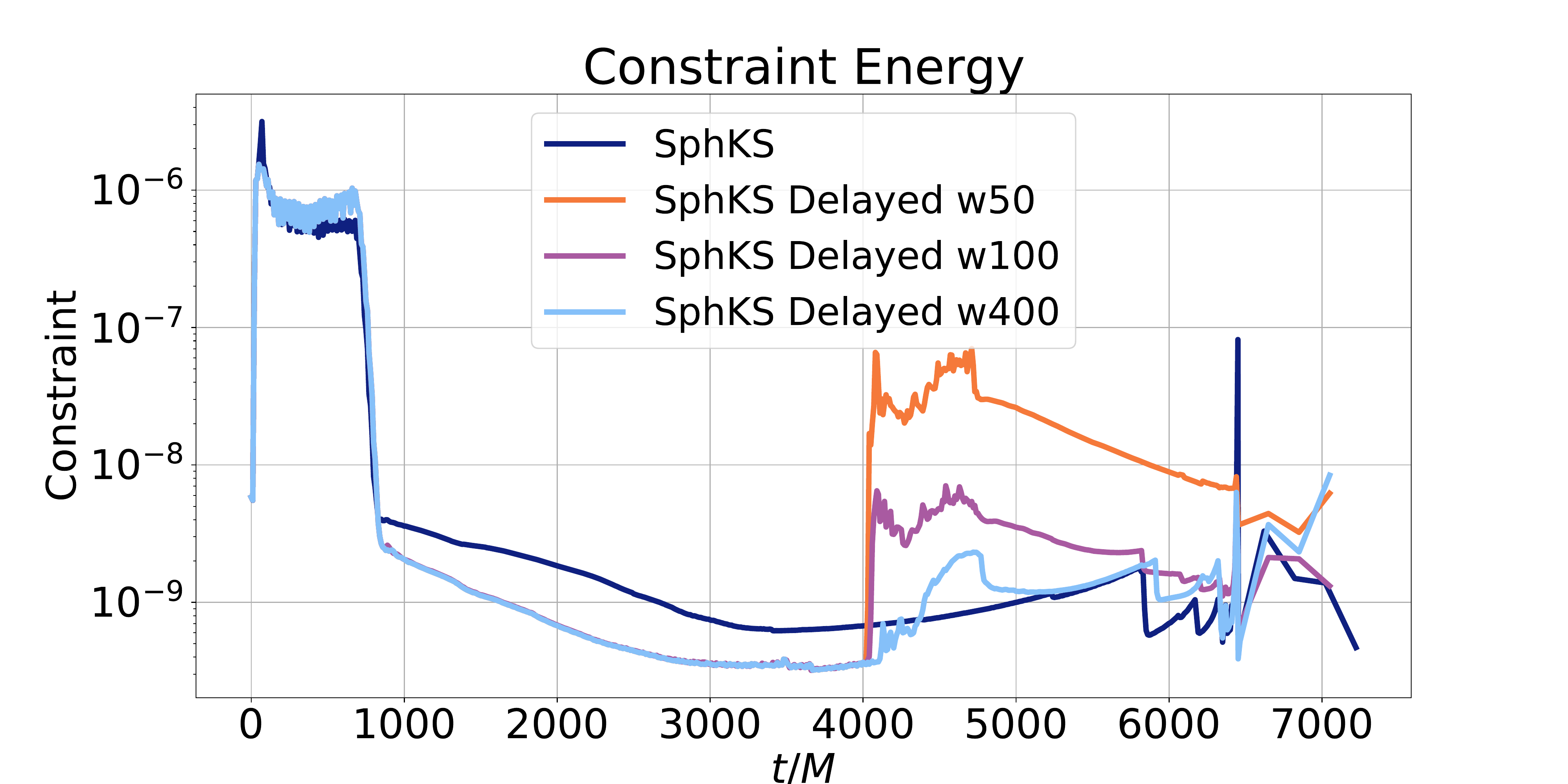}
  \caption{CPU time ratio $T/T_{\text{KS}}$ (top panel) and constraint energy
	(bottom panel) of the spin-0.9 Lev-3 BBH in SphKS and delayed SphKS. The
	speed-up by delaying the evolution gauge is manifest, by a factor of 1.3
	compared to SphKS and 1.5 compared to KS. We see that the speed-up of the
	delayed runs is mostly independent of the transition width. However, a large
	increase in the constraint violations is observed at the moment the gauge
	transition is started ($t=4000M$).  Constraint violations in the delayed gauge
	can reach several orders of magnitude higher than the non-delayed one,
	immediately after the transition time $t=4000M$.}
  \label{fig:spin0.9_bbh_delay_ghce_cpu}
\end{figure}

We plot the CPU time ratio $T/T_{\text{KS}}$ in the top panel of
Fig.~\ref{fig:spin0.9_bbh_delay_ghce_cpu}. $T_{\text{KS}}$, $T_{\text{SphKS}}$,
and $T_{\text{delay}}$ are the CPU times of simulations with the KS initial
data, the SphKS initial data without delay, and the SphKS initial data with
delayed transition of various widths $w$. A clear improvement is seen in all
delayed simulations, with the average reduction in final runtime being
$\sim25\%$ compared to SphKS and $\sim32\%$ compared to KS.

The temporal part of the DH gauge transition function has a discontinuous fourth
derivative at $t_0$ [Eq.~\eqref{eqn:time_transition}], which the high-order
methods employed by SpEC may be sensitive to. The fourth derivative is
proportional to $1/w^4$, so a narrower width introduces a larger discontinuity
in the derivative, which can lead to larger numerical errors. This feature shows
up in the constraint energy plot (the bottom panel of
Fig.~\ref{fig:spin0.9_bbh_delay_ghce_cpu}). All runs have the same order of
constraint violations before 4000$M$, but later the curves of delayed runs
abruptly jump up by several orders of magnitude at 4000$M$. The jump in the
constraint energy is larger as the temporal width gets narrower since the
derivative is larger and the gauge transition is steeper. This graph also
suggests that only a width of at least 400$M$ would be acceptable for production
BBH simulations.

\subsubsection{Waveforms}

\begin{figure}[t]
	\centering
	\includegraphics[width=\linewidth]{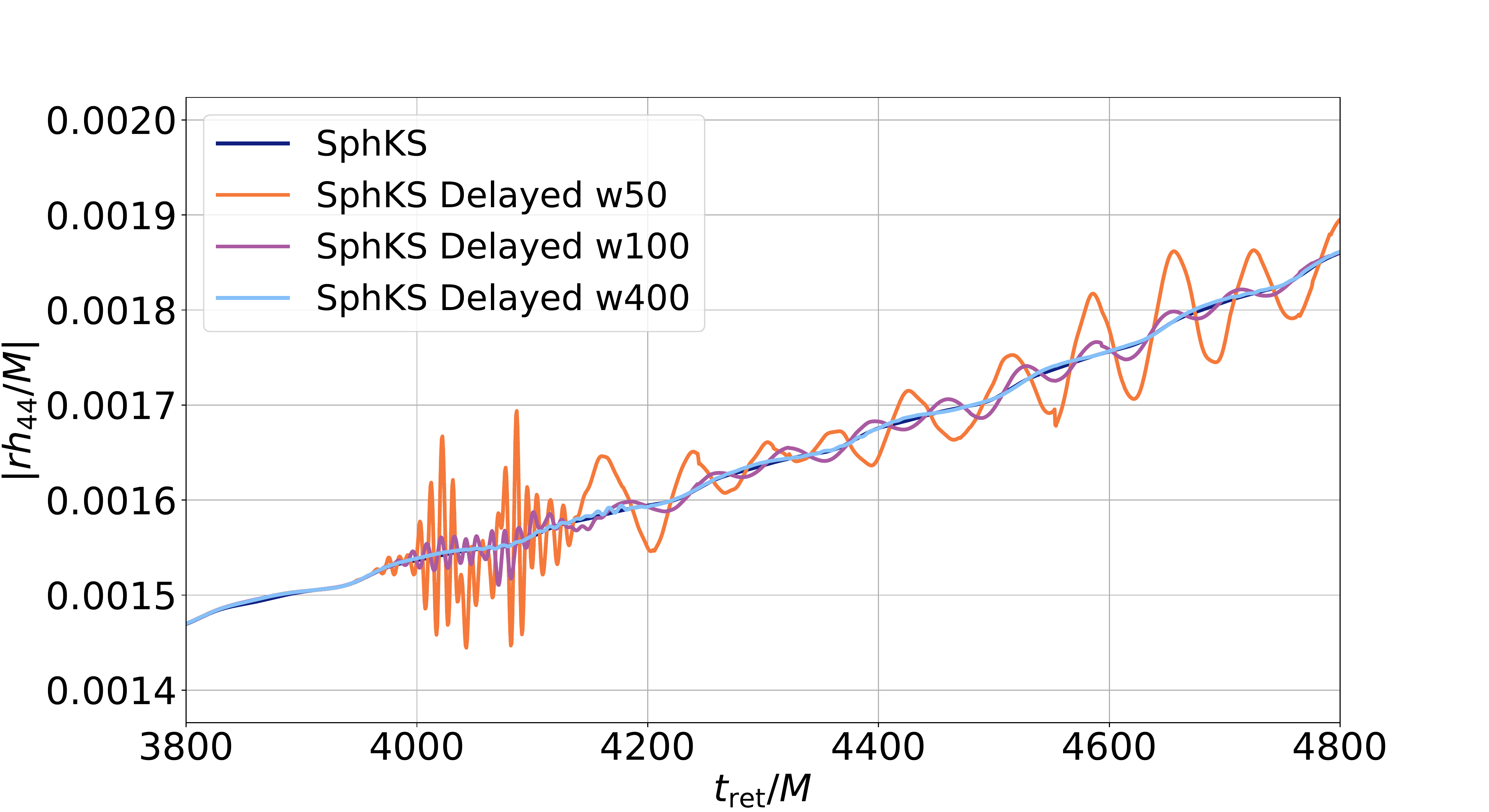}
	\includegraphics[width=\linewidth]{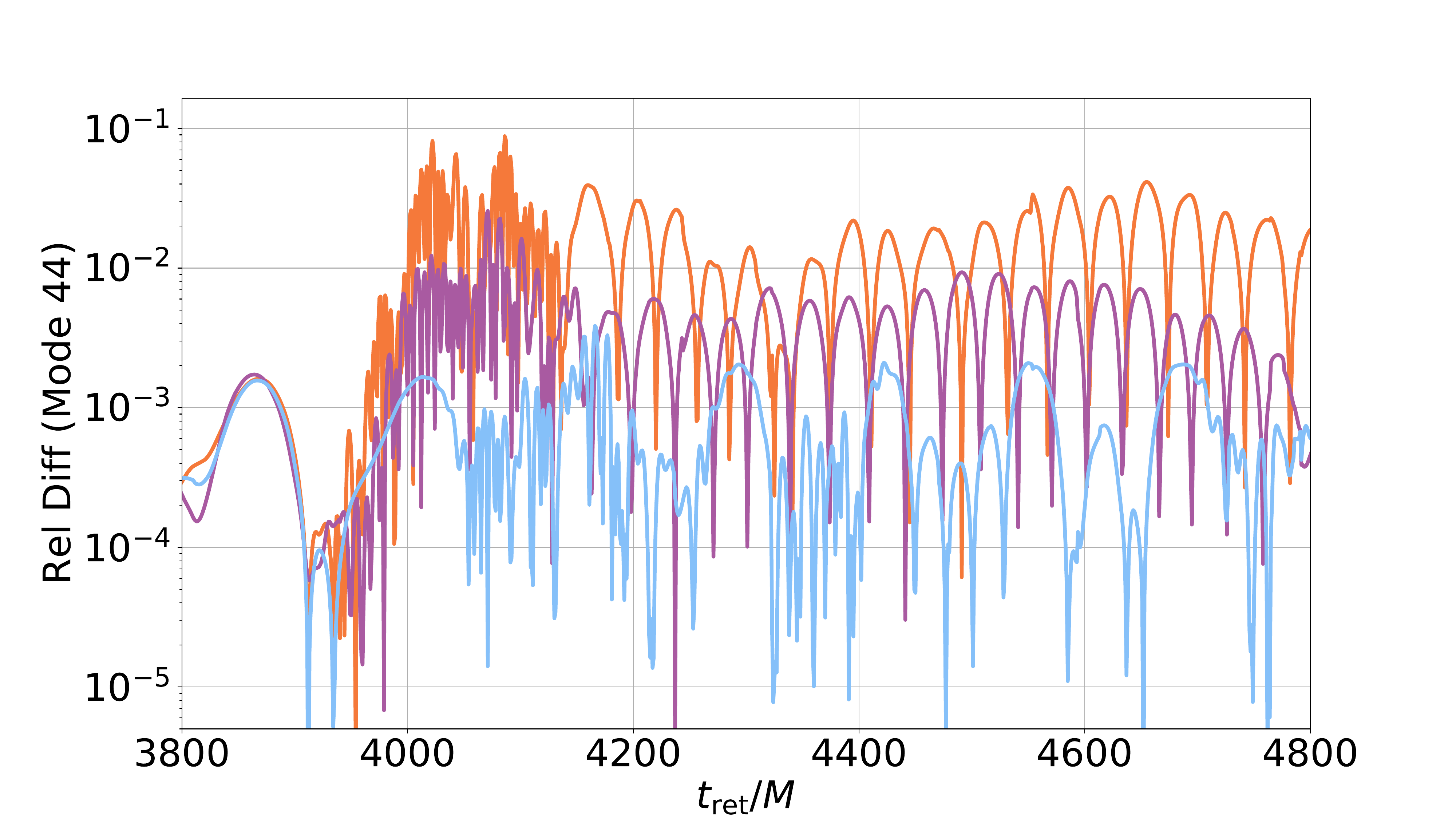}
	\caption{A comparison of the strain $rh$ between SphKS and delayed SphKS for
		the spin-0.9 Lev-3 BBH simulations. The top panel shows $|rh_{44}|$ in
		$3800M<t_{\text{ret}}<4800M$. We see the unexpected fluctuation in the
		delayed transition curves for all three temporal widths. To better see how
		the oscillations change with increasing width, the bottom panel depicts
		the relative difference in $|rh_{44}|$ of the delayed SphKS compared to the
		SphKS simulations. The fluctuation is greater as the temporal width $w$
		becomes narrower.}
	\label{fig:spin0.9_bbh_delay_waveform}
\end{figure}

The large jump in the constraint violations at $t=4000M$ may result in
unphysical effects in the waveforms. We plot $|rh_{44}|$ in the top panel of
Fig.~\ref{fig:spin0.9_bbh_delay_waveform} on the interval
$3800M<t_{\text{ret}}<4800M$. We see that large oscillations appear at $t=4000M$
in the delayed transition waveforms that are absent in the non-delayed run. The
amplitude of the oscillations decreases with increasing width. This is not
surprising considering that
the discontinuity in the fourth derivative of the roll-off
function decreases as $1/w^4$. Note that the waveforms in different gauges
overlap before the transition, suggesting that the oscillations are an artifact
of the non-smooth gauge transition.

\begin{figure*}[t]
	\centering
	\includegraphics[width=0.49\linewidth]{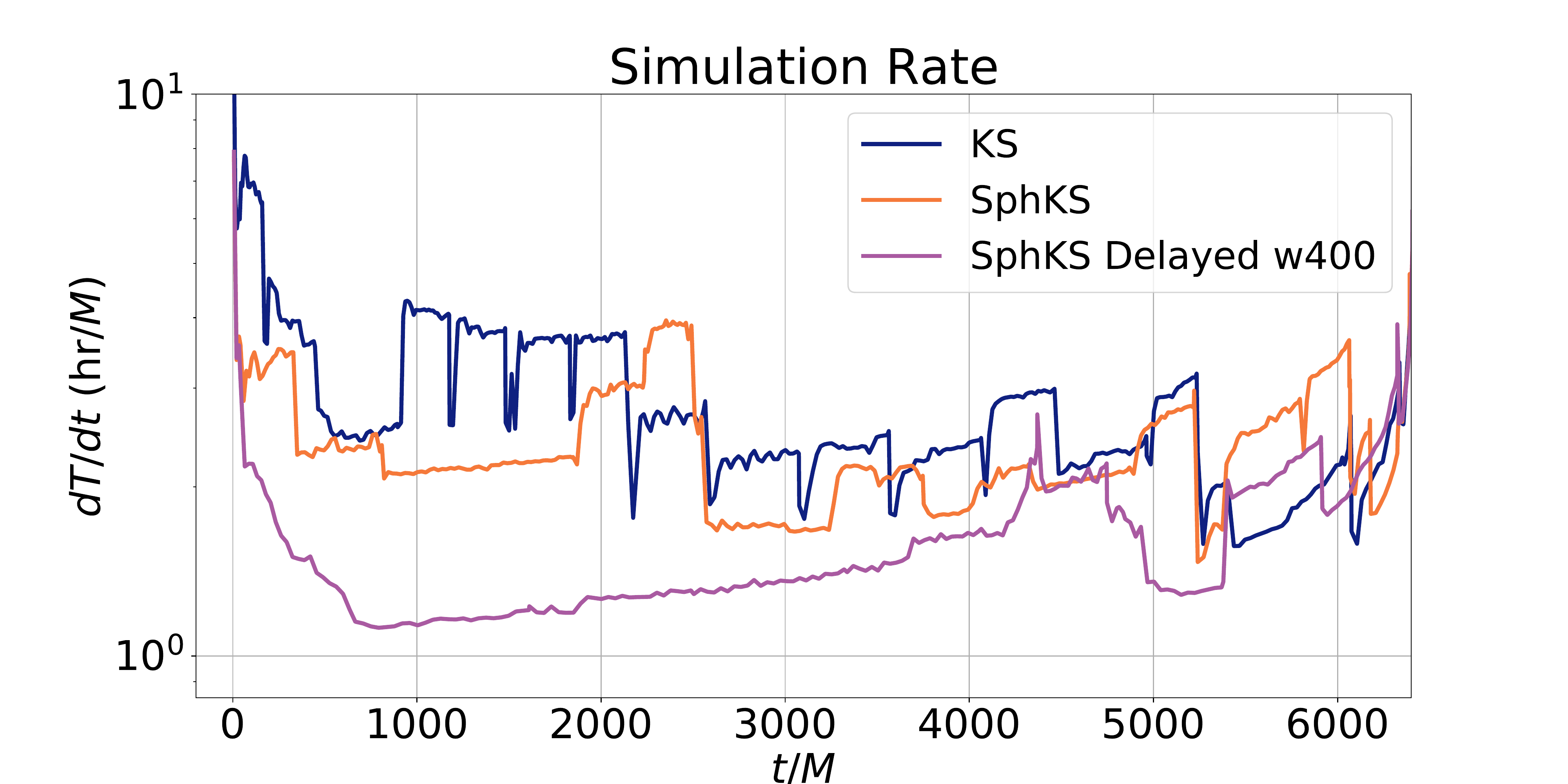}
	\includegraphics[width=0.49\linewidth]{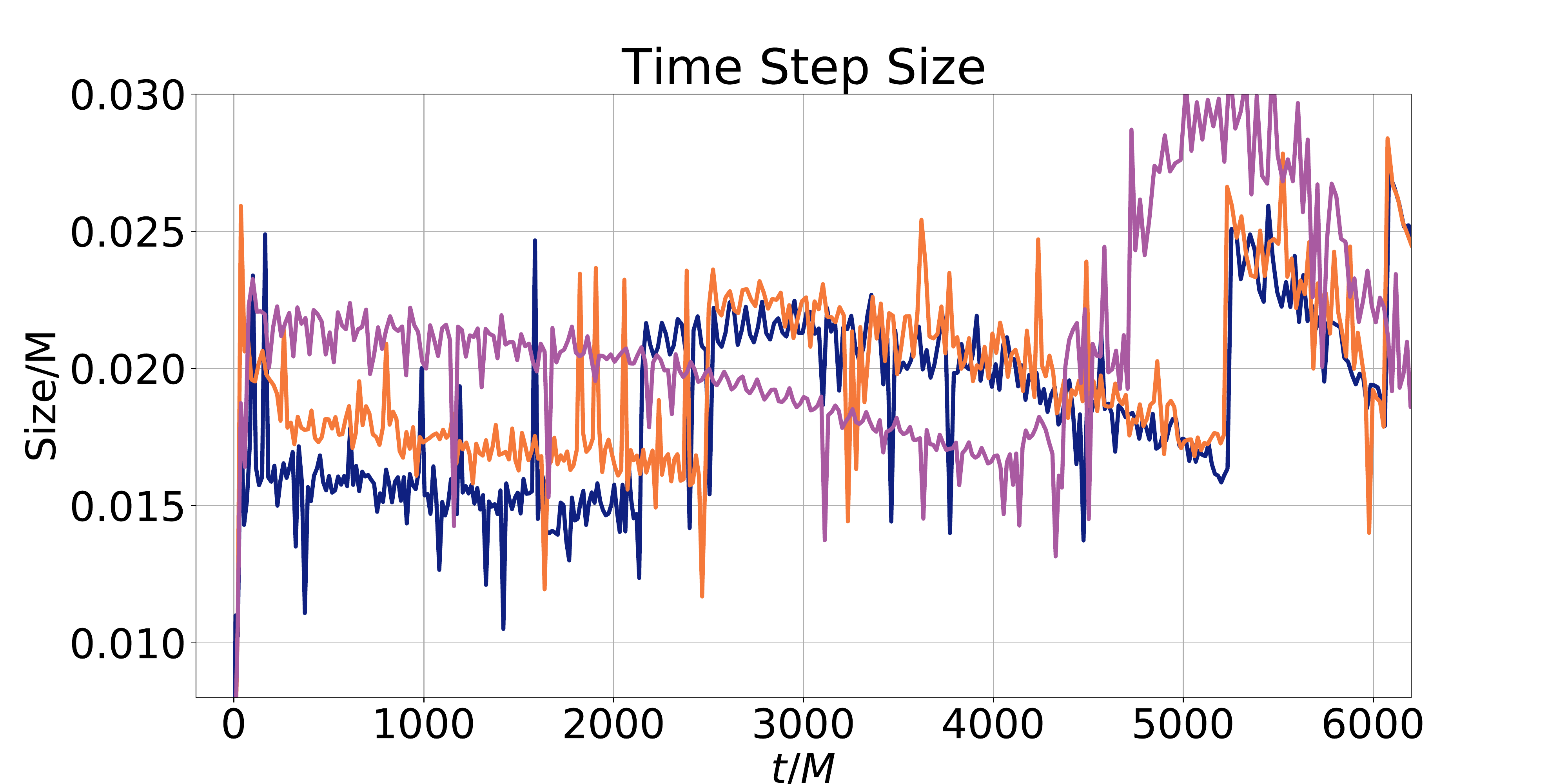}
	\includegraphics[width=0.49\linewidth]{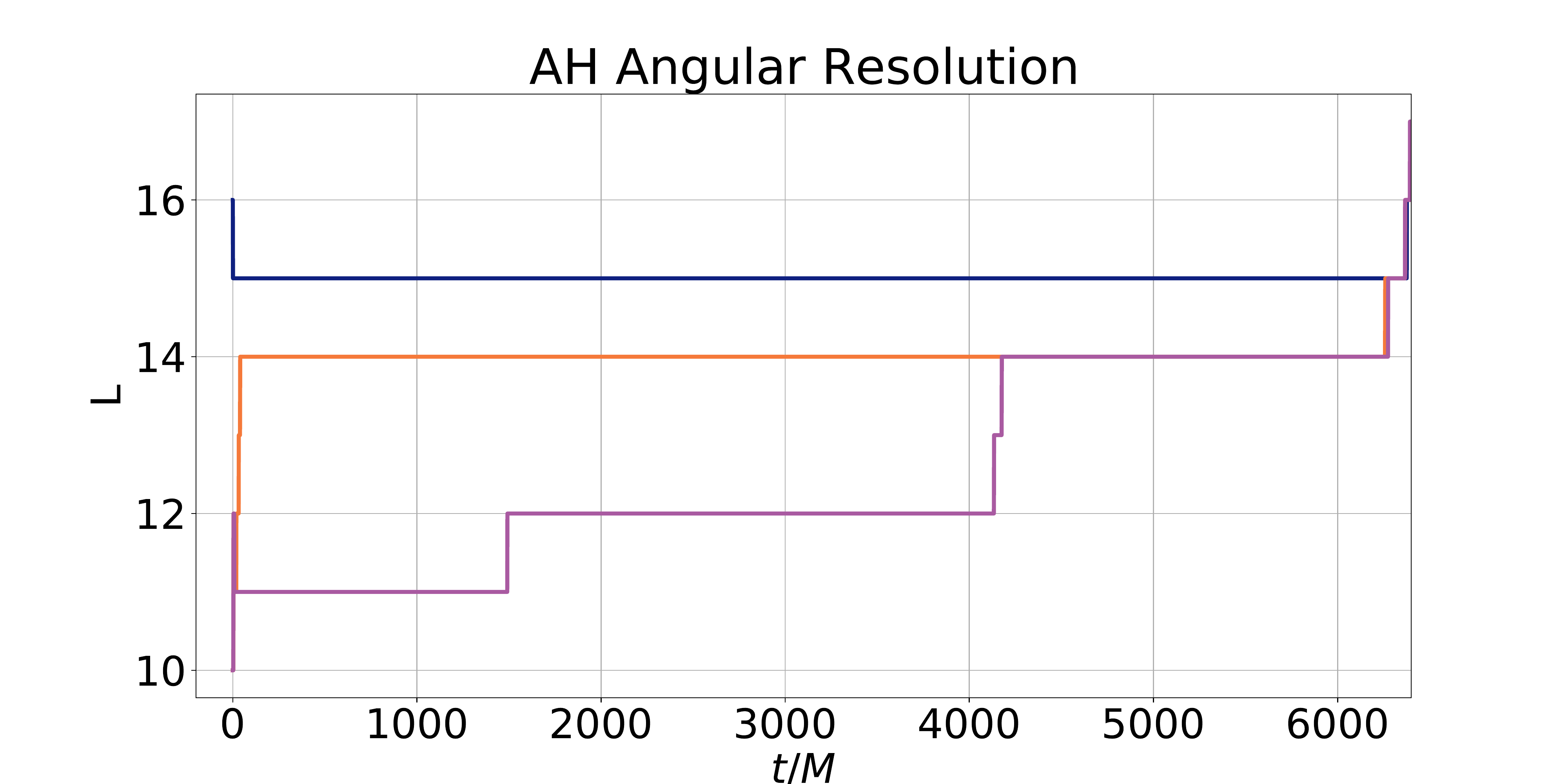}
	\includegraphics[width=0.49\linewidth]{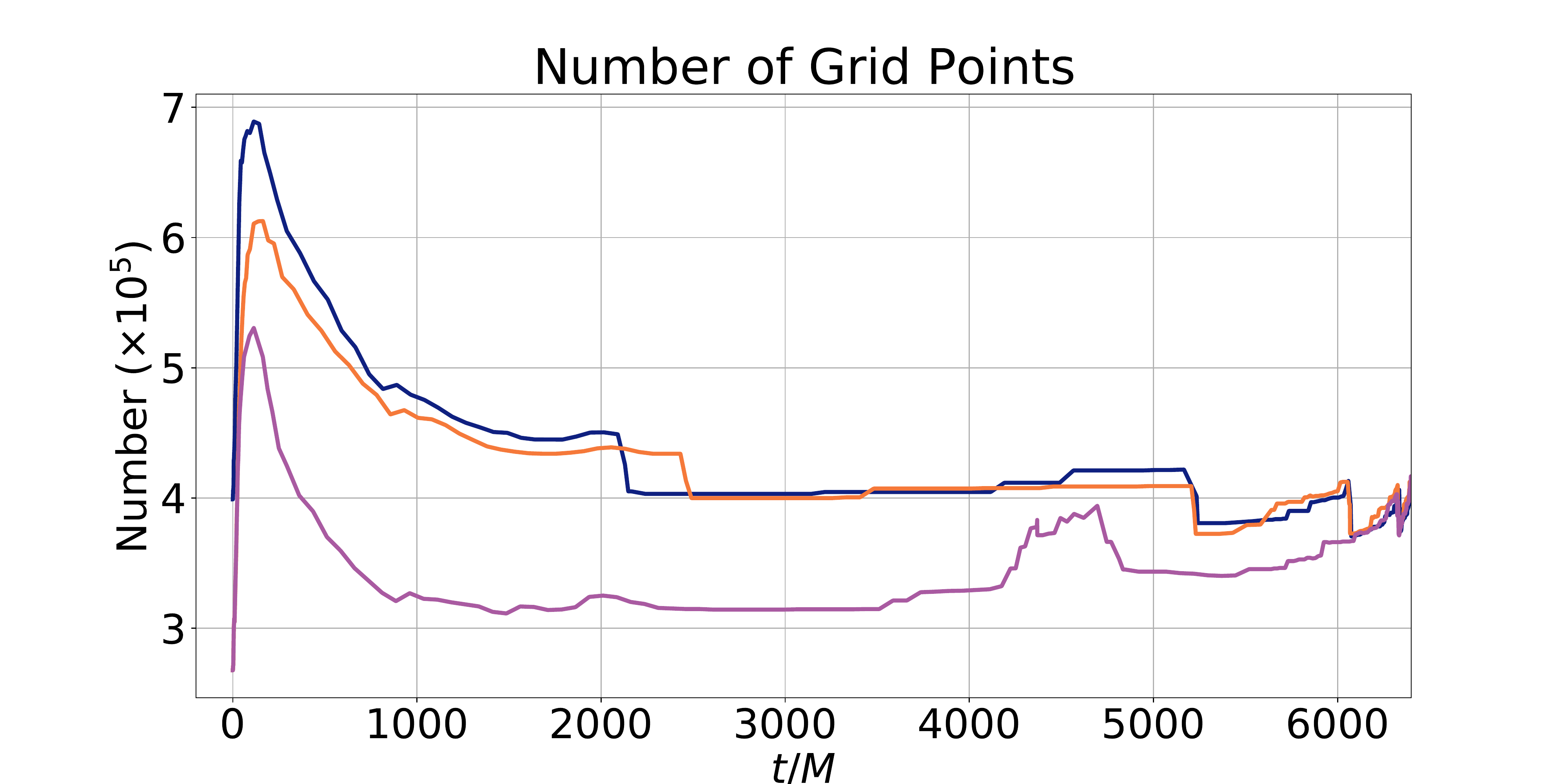}
	\caption{Four quantities for the speed-up analysis of delaying the DH gauge
		transition in the spin-0.9 Lev-3 BBH runs. The top left panel shows that a
		simulation in delayed SphKS is faster at almost every moment before 4000$M$.
		In the top right panel, we see the same level of time step sizes between
		non-delayed and delayed gauges, so time step is not a contributing factor of
		the speed-up. The bottom left graph confirms that the AH stays close to
		spherical before the onset of the gauge transition. From the bottom right
		graph, we conclude that the significantly smaller amount of grid points is
		responsible for the speed-up from delaying the gauge transition.}
	\label{fig:spin0.9_bbh_delay_speed_analysis}
\end{figure*}

To better see how the fluctuations depend on the transition width $w$, we
calculate the relative difference in $|rh_{44}|$ of the three delayed runs
compared to the non-delayed one, in the bottom panel of
Fig.~\ref{fig:spin0.9_bbh_delay_waveform}. The difference is in general smaller
as the transition width increases, which is reasonable since a narrower
transition induces a larger discontinuity in the fourth derivatives. For
example, the width-50$M$ curve has relative differences at the order of
$10^{-1}$, while the width-400$M$ differences are $10^{-3}$.

To better understand how smoothness in a delayed gauge transition function
affects a waveform, we simulate the same BBH system as above with SphKS initial
data but with a different gauge transition temporal function. Instead of
Eq.~\eqref{eqn:time_transition}, we use
\begin{align}
  F(t) =\exp\left[-\left(\frac{t}{3500M}\right)^{10}\right], \quad t\ge 0.
\end{align}
This temporal function is smooth after the start of a simulation and delays the
DH gauge until $t\sim2600M$ [when $F(t)=0.95$]. The waveforms $rh_{22}$ and
$rh_{44}$ have relative difference (compared to the non-delayed SphKS
simulation) at the same order as the $t_0=4000M, w=400M$ SphKS simulation. We find that the high-frequency noise in the waveforms is greatly reduced, but not completely eliminated. Thus, the smoothness in the temporal transition of the evolution gauge is a factor causing the high-frequency fluctuation, but not the main factor.

Given the growth in constraint violations and the appearance of fluctuations in the waveforms, we do not recommend delaying the gauge transition despite the
significant speed-up of the simulations.

\subsubsection{Speed-up analysis}

In this section, we examine the mechanism of the speed-up from delaying the DH
gauge transition.We consider three spin-0.9 Lev-3 simulations: KS initial data,
SphKS initial data without delay, and SphKS initial data with the delayed gauge
transition of width $w=400M$. Figure \ref{fig:spin0.9_bbh_delay_speed_analysis}
shows the simulation rate $dT/dt$, time step size, angular resolution $L$ used
by the AH finder, and total number of grid points in these three
simulations. The blue curves represent the simulation with the KS initial data,
the orange curves are for the non-delayed SphKS simulation, and the purple ones
for the $w=400M$ delayed transition SphKS simulation.

The graph of simulation rate (the top left panel of
Fig.~\ref{fig:spin0.9_bbh_delay_speed_analysis}) indicates that the delayed
SphKS simulation is more efficient than the other two gauges at almost any time,
especially before the transition time $t=4000M$. The top right panel of
Fig.~\ref{fig:spin0.9_bbh_delay_speed_analysis} shows that the time step sizes
for the different gauges are nearly equal, so the time step has no contribution
to the speed-up. In the bottom left panel of
Fig.~\ref{fig:spin0.9_bbh_delay_speed_analysis}, we see that the angular
resolution $L$ in the delayed SphKS simulation is initially relatively
low but then climbs to the level of the other two gauges near $t=4000M$. This is
expected because the AH is spherical in the SphKS coordinates but highly
non-spherical in the DH gauge, and delaying the transition keeps the AH in a
nearly spherical shape until $t=t_0=4000M$.  The bottom right graph of
Fig.~\ref{fig:spin0.9_bbh_delay_speed_analysis} shows that the total number of
grid points in the delayed SphKS run is considerably smaller than the other two
gauges, especially before $t=4000M$. Note that the number of grid points for
delayed SphKS is 19\%--32\% smaller than the other two gauges when
$1000M<t<4000M$, which is comparable to the overall efficiency improvement
($1-T_{\text{delay,w400}}/T_{\text{SphKS}}=26\%$). Thus, in a SphKS simulation,
delaying the DH gauge transition accelerates the computation by substantially
reducing the number of grid points.

\section{Conclusion} \label{sec:conclusion}

In this paper, we develop new gauge conditions for BHs with the goal of reducing
the computational cost of high-spin BBH simulations. We present several
different attempts, among which the most promising is the use of spherical
Kerr-Schild, where the horizons of a rotating BH are spherical. For single BH
evolutions using spherical Kerr-Schild, we find a factor of 10 reduction in the
metric error and 1000 in the constraint energy, as compared to Kerr-Schild with
the same resolution. For BBH evolutions, we see efficiency improvement with
equal accuracy. In general, we find that the speed-up is greater for simulations
with stricter truncation error tolerances and higher spin.  Specifically, we
observe an impressive factor of 2 reduction in CPU time for the
spin-0.99 Lev-3 (standard resolution of SXS BBH simulations) case. This new
gauge condition will also reduce the computational cost of extending BBH
simulations to higher spins (e.g., $\chi=0.999$), allowing waveform catalogs and
models (such as surrogates \cite{1502.07758, 1701.00550, 1705.07089, 1812.07865}) tuned to numerical relativity to cover a larger and denser
portion of the mass-spin parameter space with significantly reduced cost.

While the main focus of this paper has been improvements by changing the initial data, we also performed some experiments where we delay the transition from the initial data gauge to the damped harmonic gauge used in the evolution. The goal is to keep the horizons spherical for longer so that this further reduces the computational cost of the simulations. In Sec.~\ref{sec:spin0.9bbh_delay}, we find that imposing a spherical gauge condition during the evolution will produce an additional speed-up by a factor of 1.3. However, one must be careful not to introduce artifacts into the waveforms when delaying the gauge transition. 

Inspired by the benefit of delaying the evolution gauge, we expect a dynamical
spherical gauge condition to be very useful for simulating high-spin BBHs. As
future work, one can develop a spherical version of damped harmonic gauge, where
the horizons of BHs can remain (nearly) spherical during the whole evolution. As
far as the initial data gauge is concerned, one may consider blending the
spherical Kerr-Schild and harmonic-Kerr spatially, or even developing a
spherical version of harmonic-Kerr with the hope to reduce both the
computational cost and junk radiation. Nonetheless, solely changing the initial
data as described in this paper is certainly worthwhile.

\section*{Acknowledgments}
We thank Michael Boyle, Dante Iozzo, and Vijay Varma for useful discussions.
Computations were performed with the High Performance Computing Center and the
Wheeler cluster at Caltech. Computations were also conducted on the Frontera computing project at the Texas Advanced Computing Center. This
work was supported in part by the Sherman Fairchild Foundation and by NSF Grants
No.~PHY-2011961, No.~PHY-2011968, and No.~OAC-1931266 at Caltech, and NSF Grants
No.~PHY-1912081 and No.~OAC-1931280 at Cornell.

\appendix*
\section{$\gamma_0, \gamma_1, \gamma_2$, and $\gamma_3$ used in simulations}

SpEC evolves the spacetime of BHs in the first order
GH formalism \cite{gr-qc/0512093}, which is given by
Eqs.~(\ref{eqn:GH1}$-$\ref{eqn:GH3}). We consider four constraint
damping parameters in this formalism for this paper, namely
$\gamma_0, \gamma_1, \gamma_2$, and $\gamma_3$. These parameters
have been used to simulate BHs in previous papers. The quantities $\gamma_0$, $\gamma_1$, and $\gamma_2$ are the same
as those in Refs.~\cite{gr-qc/0512093, 1405.3693}. These three
parameters are set to nonzero values by default in a SpEC BBH
simulation. $\gamma_3$ in this paper is different from the $\gamma_3$
used in Ref.~\cite{gr-qc/0512093}. Instead, our $\gamma_3$ is the
same as the parameter $\rho$ used in Ref.~\cite{gr-qc/0504114}. The
authors of Ref.~\cite{gr-qc/0504114} set this parameter to 0 by
default, while we explore the possibility of a nonzero $\gamma_3$
for single BH simulations in this paper.

We here provide the expressions of $\gamma_0, \gamma_1, \gamma_2$,
and $\gamma_3$ used for simulations in this paper. Specifically,
for single BH simulations, we choose
\begin{align}
	\gamma_0M = \gamma_2M &= 2 \exp\left[-\left( \frac{r_O}{7M}\right)^2 \right] + 0.001, \\
	\gamma_1 &= -1, \\
	\gamma_3 &= 2,
\end{align}
where $M$ is the total ADM mass of the system as usual. $\gamma_0$ and $\gamma_2$ are spatially varying and depend on $r_O$, the Euclidean distance from the origin. We choose the origin at the geometric center of a single BH. The choice $\gamma_1 = -1$ is adopted in the simulations of Ref.~\cite{gr-qc/0512093} as well, which makes the GH system Eqs.~(\ref{eqn:GH1}$-$\ref{eqn:GH3}) linearly degenerate. Note that $\gamma_0$, $\gamma_2$ have dimension $M^{-1}$ while $\gamma_1$, $\gamma_3$ are dimensionless.

The expressions of the parameters for BBHs are more complicated than for single BHs. In the ringdown phase, we use 
\begin{align}
	\gamma_0M = \gamma_2M &= 0.001 + 7\exp\left[ -\left(\frac{r_O}{2.5M}\right)^2 \right] \nonumber \\
	&\quad + 0.1\exp\left[ -\left(\frac{r_O}{100M}\right)^2 \right],  \\
	\gamma_1 &= -1, \\
	\gamma_3 &= 0.
\end{align}
In the inspiral phase, we use
\begin{align}
	\gamma_0M = \gamma_2M &= 0.001 + 0.075 \exp\left[-\left(\frac{r_O a_\mathrm{scale}} {2.5 D_0}\right)^2\right] \nonumber \\
	&\quad+ \frac{4M}{M_A} \exp\left[-\left(\frac{r_A a_\mathrm{scale}} {7M_A}\right)^2\right] \nonumber \\
	&\quad+ \frac{4M}{M_B} \exp\left[-\left(\frac{r_B a_\mathrm{scale}} {7M_B}\right)^2\right] , \\
	\gamma_1 &= 0.999\left\{ \exp\left[ -\left(\frac{r_O}{10D_0}\right)^2 \right]-1\right\}, \\
	\gamma_3 &= 0,
\end{align}
where $M_A, M_B$ are the initial Christoudoulou masses
\cite{Christodoulou1_skip} of BH $A, B$, and $r_A, r_B$ are the
Euclidean distances from BH $A, B$. $D_0$ is the initial separation
between the two BHs, used in Tables~I and III. $a_\mathrm{scale}$
is equivalent to the (dimensionless) expansion factor $a$ used
in Refs.~\cite{gr-qc/0607056, 1211.6079}. $a_\mathrm{scale}$
is tuned by the control system in SpEC \cite{1211.6079}, so it
is time-dependent. The three distance variables $r_A$, $r_B$, and
$r_O$ are measured in the distorted frame of a BBH simulation. The
distorted frame is an intermediate frame between the grid frame
and the inertial frame, and we point interested readers
to Ref.~\cite{1211.6079} for details on the relation among these
frames. Note that we do not specify the measurement frame of $r_O$
for a single BH simulation, because the grid, the distorted and
the inertial frames are identical for single BHs in this paper.

\bibliography{library,library2}
\end{document}